\documentclass[a4paper,11pt]{article}
\pdfoutput=1 % if your are submitting a pdflatex (i.e. if you have
\usepackage{jcappub} % for details on the use of the package, please
\usepackage[utf8]{inputenc}
\usepackage{hyperref}
\usepackage{xcolor}
\usepackage{graphicx}
\usepackage{amsmath,amssymb,amsbsy,amstext,amsthm}
\usepackage{mathtools}
\usepackage{amsfonts}
\usepackage{comment}
\usepackage{bm}
\usepackage{tabularx}
\usepackage{array,collcell}

\def\be{\begin{equation}}
\def\ee{\end{equation}}

\def\fnl{{f_{\rm NL}}}
\def\fnlsq{{f^2_{\rm NL}}}
\def\OmegaGW{{\Omega_{\rm GW}}}

\def\Pk{{\mathcal{P}_{\cal R}\left(k\right)}}

\title{Probing non-Gaussianities with the high frequency tail of induced gravitational waves}
\author[a]{Vicente Atal}
\author[b]{and Guillem Dom\`enech}

\affiliation[a]{Department of Physics, University of the Basque Country, UPV/EHU, Bilbao, Spain}
\affiliation[b]{INFN Sezione di Padova, I-35131 Padova, Italy}

\emailAdd{vicente.atal@icc.ub.edu}
\emailAdd{domenech@pd.infn.it}

\abstract{We investigate in detail the spectrum of gravitational waves induced by a peaked primordial curvature power spectrum generated in single-field inflationary models. We argue that the $\fnl$ parameter can be inferred by measuring the high frequency spectral tilt of the induced gravitational waves. We also show that the intrinsically non-Gaussian impact of $\fnl$ in $\OmegaGW$ is to broaden its peak, although at a negligible level in order not to overproduce primordial black holes. We discuss possible degeneracies in the high frequency spectral tilt between $\fnl$ and a general equation of state of the universe $w$. Finally, we discuss the constraints on the amplitude, peak and slope (or equivalently, $\fnl$) of the primordial power spectrum by combining current and future gravitational wave experiments with limits on $\mu$ distortions from the cosmic microwave background.}

\begin{document}

\maketitle

\section{Introduction}

Gravitational waves (GW) are a very interesting probe to gain knowledge of the history and content of the Universe. The large number of present and future experiments in the search of GW signals, such as LISA \cite{Audley:2017drz}, Taiji \cite{Guo:2018npi}, Tianqin \cite{Luo:2015ght}, DECIGO \cite{Seto:2001qf,Yagi:2011wg}, AION/MAGIS \cite{Badurina:2019hst}, ET \cite{Maggiore:2019uih} and PTA \cite{Lentati:2015qwp}, ensures a large improvement in our understanding of the different periods of the cosmological history. One of the most exciting possibilities is obtaining new information about the early period of cosmological inflation \cite{Brout:1977ix,Starobinsky:1980te,Guth:1980zm,Sato:1980yn}. There is at present little constraints on the energy scale of inflation. Furthermore, the constraints to the inflationary potential are limited to a small range of scales. In particular, while the study of the Cosmic Microwave Background (CMB) by the Planck team provided an extremely accurate measurement of the primordial spectrum on the largest scales \cite{Akrami:2018odb,Aghanim:2018eyx}, which corresponds to the first few observable e-folds of inflation, the nature of the power spectrum for smaller scales is poorly constrained. While an extrapolation of the physics of the simplest models of inflation beyond CMB scales would render a GW detection very difficult, it might be possible that perturbations at small scales are actually enhanced with respect to CMB scales. 

This possibility is of great interest since an enhancement of the power spectrum could result in copious productions of Primordial Black Holes (PBHs), which could play a role explaining dark matter \cite{Hawking:1971ei,Carr:1974nx,Carr:2016drx,Inomata:2017okj}. See Ref.~\cite{Sasaki:2018dmp} for a recent review. Even if PBH do not constitute all of the dark matter, binaries of PBH might account for some of the LIGO/VIRGO gravitational waves (GW) events \cite{Bird:2016dcv,Clesse:2016vqa,Sasaki:2016jop,Atal:2020igj,Wong:2020yig} and the microlensing events by planet-mass objects found by OGLE \cite{2017Natur.548..183M,Niikura:2019kqi}.
PBH might also be the seeds of supermassive black holes \cite{Kawasaki:2012kn,Carr:2018rid}.

Now, if the power spectrum of scalar perturbations is enhanced with respect to CMB scales, the resulting secondary GW background signal could be observed by ongoing or future experiments \cite{tomita,Matarrese:1992rp,Matarrese:1993zf,Matarrese:1997ay,Carbone:2004iv,Ananda:2006af,Baumann:2007zm,Saito:2008jc}. The so-called induced GWs have received a lot of attention recently \cite{Alabidi:2012ex,Alabidi:2013wtp,Garcia-Bellido:2017aan,Inomata:2016rbd,Orlofsky:2016vbd,Domenech:2017ems,Espinosa:2018eve,Kohri:2018awv,Cai:2018dig,Bartolo:2018rku,Inomata:2018epa,Unal:2018yaa,Clesse:2018ogk,Yuan:2019udt,Inomata:2019zqy,Inomata:2019ivs,Chen:2019xse,Domenech:2019quo,Ota:2020vfn,Cai:2019jah,Yuan:2019wwo,Cai:2019elf,Cai:2019amo,Bhattacharya:2019bvk,Pi:2020otn,Xu:2019bdp,Yuan:2020iwf,Ballesteros:2020qam,Liu:2020oqe,Ozsoy:2020ccy,Ozsoy:2020kat,Braglia:2020eai,Braglia:2020taf,Fumagalli:2020nvq,Riccardi:2021rlf}. They are a crucial counterpart of PBHs as a fraction, if not all, of dark matter \cite{Espinosa:2018eve,Cai:2018dig,Bartolo:2018rku,Yuan:2019udt} and their absence severely constrains the PBH reheating scenario \cite{Papanikolaou:2020qtd,Domenech:2020ssp}. Induced GWs also constitute a powerful probe of the primordial curvature power spectrum \cite{Inomata:2018epa,Gow:2020bzo} 
and of the thermal history of the universe \cite{Cai:2019cdl,Hajkarim:2019nbx,Domenech:2019quo,Domenech:2020kqm}. Furthermore, they are a candidate to the recent observation by NANOGrav \cite{Bian:2020bps,Arzoumanian:2020vkk} of a possible stochastic GW background \cite{Vaskonen:2020lbd,DeLuca:2020agl,Kohri:2020qqd,Vagnozzi:2020gtf,Domenech:2020ers,Inomata:2020xad,Bhattacharya:2020lhc}. It should also be noted that the issue of the seeming gauge dependence of the induced GW spectrum \cite{Hwang:2017oxa,Gong:2019mui,Tomikawa:2019tvi,Yuan:2019fwv} has been settled to a great extent \cite{DeLuca:2019ufz,Inomata:2019yww,Domenech:2020xin}.\\

In this work we give a precise estimation of the spectral features of the induced GW background, $\OmegaGW$, as coming from single-field inflation, taking into account the shape of the power spectrum of scalar perturbations and the non-Gaussian nature of the perturbations. In particular we show that:

i) The spectral index of the induced GWs after the peak is directly related to the parameter $\fnl$ that determines the size of the primordial non-Gaussianities. We show this relation in general for different dominant component of the equation of state of the Universe at the moment that the perturbations re-enter the horizon.  

ii) The spectral features of the induced GWs are robust against the non-Gaussian corrections to the 4-point function of the scalar perturbations sourcing them. While these have the effect of broadening the peak of $\OmegaGW$, the effect is negligible for all the relevant range of parameters. 
From these observations we perform an analysis on how current and future observations constraint the parameters of the power spectrum as coming from single-field inflationary models.

In single-field inflation models, the spectrum of fluctuations featuring a peak corresponds to a broken power-law \cite{Atal:2018neu}. The induced GWs coming from such power spectra has been considered in Refs. \cite{Clesse:2018ogk,Xu:2019bdp,Liu:2020oqe,Riccardi:2021rlf}. Here we go beyond these works showing estimates for both the amplitude and spectral tilt of the induced GWs, by generalizing them to arbitrary cosmological backgrounds, by computing their exact non-Gaussian corrections, and by showing the largest possible impact of the latter by virtue of the relations existing between $\fnl$, the threshold for critical collapse and the abundance of PBHs.

This work is organized as follows. In Sec.~\ref{sec:induced} we review the enhancement of primordial fluctuations in single-field inflationary models. We emphasize the relation between the ultraviolet spectral tilt and the magnitude of non-Gaussianities. We also provide accurate analytical formulas for the GW spectrum induced by a broken power-law primordial spectrum. In Sec.~\ref{sec:nonGaussianities} we investigate in detail the impact of non-Gaussianities, inevitably generated during the enhancement of the primordial fluctuations. Later in Sec.~\ref{sec:observations} we consider how current and future observations might constrain the model. We conclude our work with discussions in Sec.~\ref{sec:conclusions}. Details of the calculations and useful formulas can be found in the appendices. In particular, in App.~\ref{app:calculations} we derive general compact formulas of the induced GWs including non-Gaussianities. In App.~\ref{app:walnut} we give analytical approximations for the non-Gaussian contribution to the GWs induced by a broken power-law primordial spectrum, including the contribution from the connected 4 point function.
 
\section{The induced Gravitational Waves in Single-Field Inflation \label{sec:induced}}

In this section we provide analytical formulas for the GW spectrum induced by a peak in the curvature power spectrum arising in general single-field models of inflation. We first review the most relevant aspects of the resulting shape of the primordial curvature power spectrum which are key in determining $\OmegaGW$. Namely, we focus on the UV spectral tilt and the amplitude and shape of the non-Gaussianities, and how these are related to the local potential. We then turn to the calculations of the induced GW background.

\subsection{The shape of the primordial power spectrum and its relation to \texorpdfstring{$\fnl$}{fnl} \label{subsec:primordialspectrum}}

In single-field inflation, it is possible to generate a large enhancement of the scalar perturbations with respect to CMB scales if the inflaton traverses a local maximum of the potential. We may model the resulting power spectrum as a broken power-law, namely
\begin{align}\label{eq:PR}
\Pk={\cal A}_{\cal R}\left\{
\begin{aligned}
&\left(\frac{k}{k_p}\right)^{n_{\rm IR}}\quad &k\leq k_p\\
&\left(\frac{k}{k_p}\right)^{-n_{\rm UV}}\quad &k\geq k_p
\end{aligned}
\right.\,,
\end{align}
where ${\cal A}_{\cal R}$ is the amplitude of the power spectrum at the peak, $k_p$ is the position of the peak and $n_{\rm IR}$ and $n_{\rm UV}$ are positive constants respectively referred to as the spectral tilt in the infrared (IR) limit ($k\ll k_p$) and in the ultraviolet (UV) limit ($k\gg k_p$).
The IR limit corresponds to scales that exit the horizon during the period  in which the inflaton field is climbing the local maximum of the potential. In this stage there is a phase of ultra-slow roll and the scalar perturbations grow after crossing the horizon. Determining the spectral index is rather complicated but can be done analytically.  It has been found that at the IR scales \cite{Byrnes:2018txb,Ozsoy:2019lyy}
\be\label{eq:nir_4}
n_{\rm IR}\lesssim 4 \ .
\ee
While steeper spectra can be reached for some specific backgrounds in single-field \cite{Carrilho:2019oqg,Tasinato:2020vdk}, or in multi-field scenarios \cite{Palma:2020ejf,Fumagalli:2020adf,Braglia:2020taf}, Eq. \eqref{eq:nir_4} describes well the IR slope in the most minimalistic models resulting in a peak in single-field. Thus, in this work we take $n_{\rm IR}\sim 4$, although our formulas can be easily applied to steeper slopes.

After reaching the maximum of the potential, the field rolls down a possibly very steep potential. After few efoldings, there is no superhorizon growth of the perturbations. Since the velocity of the field is exponentially suppressed, the spectral index is dominated by the second slow-roll parameter, and so
\be
n_{\rm UV}=\epsilon_2 \ .
\ee
The amplitude and shape of non-Gaussianities around the peak can be retrieved from the structure of the potential around the local maximum. That is, we can consider the potential as 
\be\label{eq:potential}
V= \frac{\eta_V}{2}\left(\phi-\phi_o\right)^2\,,
\ee
where $\eta_V < 0$. Considering this potential, curvature perturbations follow a local non-Gaussian distribution described by \cite{Atal:2019cdz}
\be\label{eq:ng_np}
{\cal R}=-\frac{2}{\epsilon_2^{ge}}\ln\left(1-\frac{\epsilon_2^{ge}{\cal R}_G}{2}\right)\,,
\ee
where $\epsilon_2^{ge}\equiv-3+\sqrt{9-\eta_V} $ is the second slow roll parameter after crossing the local maxima (corresponding to the  \textit{graceful exit} from USR). For small ${\cal R}_G$, Eq. \eqref{eq:ng_np} can be expanded and we retrieve the known quadratic formulation of local non-Gaussianity.
\be\label{eq:ng_pert}
{\cal R}={\cal R}_G+\frac{3}{5}\fnl{\cal R}_G^2\,,
\ee
where $\fnl$ is given by
\be\label{eq:fnl_epsilon2}
\fnl=\frac{5}{12}\epsilon_2^{ge}\,.
\ee
From here we deduce the following simple relation, already noticed in \cite{Atal:2019erb} (see also \cite{Taoso:2021uvl})
\be\label{eq:nuvfnl}
n_{\rm UV}=\frac{12}{5}\fnl\,.
\ee
Let us note that $\fnl=\frac{5}{12}n_{\rm UV}$ for all scales that exit the horizon during the time in which the potential is given by (\ref{eq:potential}). This means that $\fnl$ is constant and possibly large for a range of scales comprising not only the peak of the power spectrum, but also an important part of the rising and decaying tails around it \cite{Atal:2018neu}\footnote{For recent discussions on the observability of $\fnl$ in the squeezed limit, see Refs. \cite{Matarrese:2020why,Bravo:2020hde,Suyama:2021adn}.}.

A perturbative estimation of non-Gaussianities can also be found for other shapes of the potential \cite{Cai:2017bxr,Suyama:2021adn}, although the simple quadratic potential presented here is sufficient for accurately estimating $\fnl$ for most of the models present in the literature \cite{Atal:2018neu}. The only regime in which it does not accurately predict the amplitude of $\fnl$ (for realistic smooth potential) is when the potential is exactly flat in the USR phase. Non-Gaussianities are still perturbatively local, as determined by Eqs. (\ref{eq:ng_pert})-(\ref{eq:fnl_epsilon2}), but $\epsilon_2^{ge}$ will now depend differently on the potential (since $\eta_V=0$). In this case however $\fnl\ll1$. Thus, in the most interesting regime for non-Gaussianities, $\fnl\gtrsim 0.1 $, our model describes all smooth single-field models.

In the following we proceed to estimate the induced GWs from a broken power-law power spectrum \eqref{eq:PR}. 

\subsection{Gravitational waves induced by a broken power-law \label{subsec:GWpowerlaw}}

In this section we provide analytical estimates for the infrared (IR) and ultraviolet (UV) tails of the induced GW spectrum assuming that the primordial scalar spectrum is given by the broken power-law in Eq.~\eqref{eq:PR}. We use such estimates to constraint the magnitude of the local non-Gaussianity associated to peaked scalar spectrum generated in single-field inflationary models. The basic formula for the induced GW spectral density for a given wavenumber $k$ can be found, e.g., in Refs.~\cite{Kohri:2018awv,Domenech:2019quo} and it is given by
\begin{align}\label{eq:OGW}
\Omega_{\rm GW}(k,\tau)=\frac{k^2}{12{\cal H}^2}{\cal P}_h(k,\tau)\,,
\end{align}
where $\tau$ is the conformal time, ${\cal H}$ is the conformal Hubble parameter and ${\cal P}_h(k,\tau)$ is the induced tensor mode (dimensionless) power spectrum summed over the two polarizations, which explicitly reads
\begin{align}\label{eq:ph}
{\cal P}_h(k,\tau)=8\int_0^\infty dv\int_{|1-v|}^{1+v}du\left(\frac{4v^2-(1-u^2+v^2)^2}{4uv}\right)^2\overline{I^2(\tau,k,v,u)}{{\cal P}_{\cal R}(ku)}{{\cal P}_{\cal R}(kv)}\,.
\end{align}
In Eq.~\eqref{eq:ph}, ${\cal P}_{\cal R}(k)$ is a general (dimensionless) primordial curvature power spectrum, $I(\tau,k,v,u)$ is the kernel and an overline denotes oscillation average. Details of the derivation and formulas can be found in App.~\ref{app:calculations}. Despite the compact form of the tensor mode power spectrum \eqref{eq:ph}, it is technically difficult to obtain general analytical formulas due to the non-trivial form of the kernel. In particular, the averaged kernel squared in a radiation dominated universe is given by
\begin{align}\label{eq:kernelRD}
\overline{I_{RD}^2(u,v,x\gg1)}=&\frac{1}{2}\left(\frac{3y}{2uvx}\right)^2\left(\pi^2y^2\Theta(u+v-\sqrt{3})+\left(2-y\ln\left|\frac{1+y}{1-y}\right|\right)^2\right)\,.
\end{align}
where we introduced
\begin{align}\label{eq:y}
y\equiv\frac{u^2+v^2-3}{2uv}\,,
\end{align}
for an easier later comparison with other cosmological backgrounds. See App.~\ref{app:kernelgeneral} for the analytical expressions for an arbitrary equation of state and sound speed of scalar fluctuations.

As argued in Sec.~\ref{subsec:primordialspectrum}, the enhanced primordial curvature spectrum generated during single-field inflation has an approximate broken power-law shape given by Eq.~\eqref{eq:PR}. By glancing at Eq.~\eqref{eq:ph} one is tempted to say that the $k$ dependence of the tensor spectrum is roughly ${\cal P}_h(k)\sim {\cal P}^2_{\cal R}(k)$. However, this is not true in general. For instance, it is known that for localized sources the IR tail of the GW spectrum in general goes as $k^3$ \cite{Cai:2019cdl}. As an illustrative example, let us take a single power-law ${\cal P}_{\cal R}(k)\sim k^\sigma$ for all wavenumbers and analyze the convergence of Eq.~\eqref{eq:ph}. First of all, by inspecting the kernel \eqref{eq:kernelRD} we see that $\overline{I^2}$ decays at the boundaries of the $u,v$ plane. We note that there is a divergence at $u+v=\sqrt{3}$, where $y=-1$ and the logarithm blows up. Nevertheless, such divergence has zero measure and it is irrelevant under the integral. Thus, whether the integral converges depends on the asymptotic behavior of ${\cal P}_{\cal R}(k)$ at either $k\to \infty$ or $k\to 0$. We find that the integral \eqref{eq:ph} converges if $-4<\sigma<3/2$. The details of the calculations can be inferred from the discussion below.

\begin{figure}
\centering
\includegraphics[width=0.6\columnwidth]{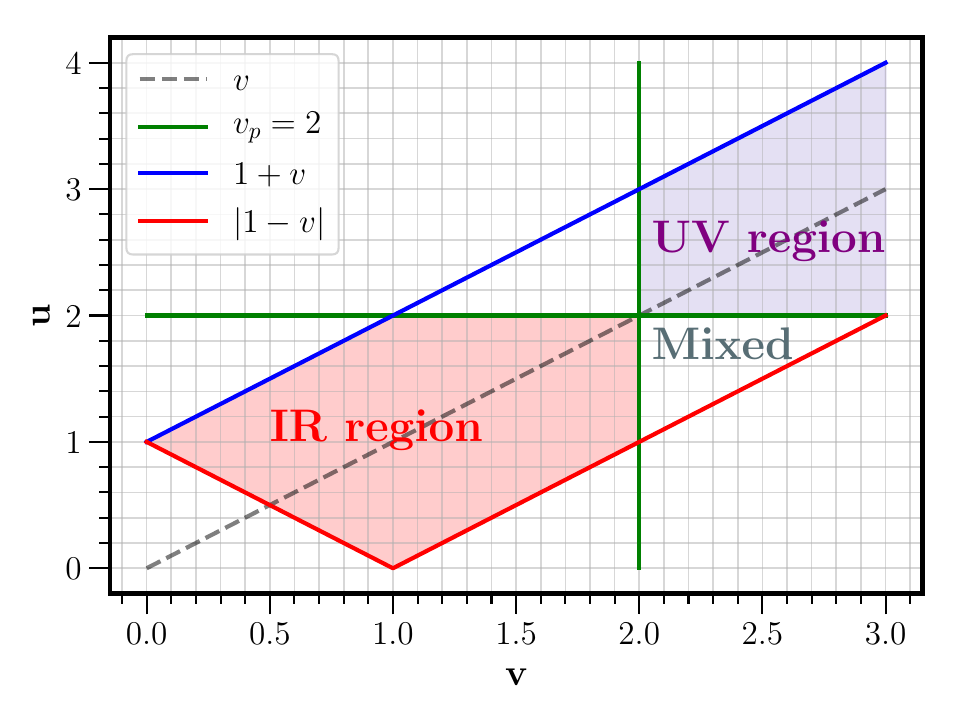}
\caption{Integration plane of Eq.~\eqref{eq:ph} in terms of the $v,u$ variables. The blue and red lines respectively correspond to the limits $u=1+v$ and $u=|1-v|$. We treat the broken power-law spectrum strictly as two power-laws: the IR power-law with a UV cut-off at $k=k_p$ and the UV power-law with an IR cut-off at $k=k_p$. The green line depicts a given cut-off for $u$ and $v$ at $k_p=2k$, that is at $u_p=k_p/k=2$ and $v_p=k_p/k=2$. Due to the presence of the cut-offs we can distinguish 4 zones in the integration plane. When in the integrand we have the two IR (UV) power-laws we call it the IR (UV) region highlighted in red (purple). Otherwise, when we have one IR and one UV power-law, we call it the mixed region. The IR limit of the GW spectrum corresponds to sending the green lines to infinity as $k\ll k_p$, or $v_p\gg1$. The opposite case, $v_p\ll1$, corresponds to the UV limit of the GW spectrum. \label{Fig:regions}}
\end{figure}

For the case at hand, the power spectrum \eqref{eq:PR} is well behaved in both limits. However, we may treat Eq.~\eqref{eq:PR} as two independent power-laws with cut-offs at $k=k_p$ and calculate their contribution to the GW spectrum separately. This is clear from Fig.~\ref{Fig:regions}. Thus, in practice, the integral in each region of Fig.~\ref{Fig:regions} effectively sees a single power-law and the previous discussion on the convergence becomes useful. Since the peak in the power spectrum inside the integrand occurs at $v_p=k_p/k$ (the green lines of Fig.~\ref{Fig:regions}), we have that in the IR limit ($v_p\gg1$) the IR region of integration dominates. In the UV limit ($v_p\ll 1$), we have that the UV region of integration dominates. We checked that the contribution from the other regions is often subleading, unless the UV spectral tilt is small. It is also interesting to notice that the regions of integration and the integrand are symmetric with respect to the $u=v$ line, which simplifies some calculations.

We first consider the IR limit of the GW spectrum. In terms of the UV cut-off this corresponds to $v_p\gg 1$ and the power spectrum is to a good approximation ${\cal P}_{\cal R}(k)\approx {\cal A}_{\cal R} (k/k_p)^{n_{\rm IR}}$. Since the power-spectrum grows with $k$, the dominant contribution is at large momenta. In terms of the variables in Eq.~\eqref{eq:ph}, this corresponds to the limit $u\sim v\gg1$ and $y\sim1$. In this limit, the kernel decays as $\overline{I^2}\sim v^{-4}$, where we neglected the logarithm term for simplicity. Thus, the integrand in Eq.~\eqref{eq:ph} goes as $v^{2 n_{\rm IR}-4}$. Now, the integral over $u$ can be approximated\footnote{A more accurate calculation involves introducing two new variables, $t=u+v$ and $s=u-v$, and expand for large $t$. To a good approximation the integral over $s$ may be evaluated at $s=0$, with an error of less than $O(1)$.} by the integrand evaluated at $u=v$, since the range of $u$ is infinitely small for $v\to\infty$. For $n_{\rm IR}>3/2$, which is the current case of study, the dominant contribution lies at the cut-off $v_p$. Taking into account the numerical factors, expanding the integrand of \eqref{eq:ph} for $u\sim v\gg 1$, we find that Eq.~\eqref{eq:OGW} in the IR limit is given by
\begin{align}\label{eq:OGWIR}
\Omega^{\rm IR}_{\rm GW}(k\ll k_{p})\approx 3\left(\frac{1}{2n_{\rm IR}-3}+\frac{1}{2n_{\rm UV}+3}\right){\cal A}_{\cal R}^2\left(\frac{k}{k_p}\right)^{3}\ln^2\left(\frac{k}{k_p}\right)\,,
\end{align}
where we have included the contribution from the UV region for better accuracy when $n_{\rm UV}$ is small.
This is the well-known result that the GW spectrum for localized sources in a radiation dominated universe in the IR limit goes as $k^3$ \cite{Cai:2019cdl}. The logarithmic running is characteristic of the induced GWs in a radiation dominated universe \cite{Yuan:2019wwo}. See Ref.~\cite{Domenech:2020kqm} for the values of the equation of state parameter $w$ for which the logarithmic running is present/absent.  Since we are mainly interested in $n_{\rm IR}\sim 4$, Eq.~\eqref{eq:OGWIR} gives a good order of magnitude estimate. 

Let us turn our attention to the UV limit of the GW spectrum. This time we have that the cut-off is an IR cut-off at $v_p\ll 1$ with ${\cal P}_{\cal R}(k)\approx {\cal A}_{\cal R} (k/k_p)^{-n_{\rm UV}}$. In terms of $u$ and $v$ the UV limit corresponds to $(u\to 1,v\to 0)$ and $(u\to 0,v\to 1)$. The latter regime is a copy of the former due to the symmetry of the integrand along the $u=v$ line. Focusing only in the former we see that the integral over $u$ for $v\ll1$ is well approximated by twice the integrand evaluated at $u=1$ times $v$. The kernel in this limit goes as $\overline{I^2}\sim \rm constant$ as $y\sim -1/v$ and so the integrand is proportional to $v^{3-n_{\rm UV}}$. Note that the integral converges if $n_{\rm UV}<4$ in the strict limit when $v_p\to 0$. Thus, in the analytical estimates of the UV tail of the induced GW spectrum we have to treat the cases $n_{\rm UV}<4$ and $n_{\rm UV}>4$ separately.

On one hand, when $n_{\rm UV}<4$ the integral converges everywhere in the plane $u,v$ even when $v_p\to 0$. This implies that most of the contribution comes from the divergent part of the kernel. We could use similar approximations as in Ref.~\cite{Inomata:2019ivs} and restrict ourselves to the close neighborhood of the divergence. However, it is more practical to write the spectrum as
\begin{align}\label{eq:OGWUV}
\Omega^{\rm UV}_{\rm GW}(k\gg k_{p},n_{\rm UV}<4)\approx \frac{1}{12}{\cal A}_{\cal R}^2F(n_{\rm UV})\left(\frac{k}{k_p}\right)^{-2n_{\rm UV}}\,,
\end{align}
where $F(n_{\rm UV})$ is given by
\begin{align}
F(n_{\rm UV})=8\int_0^\infty dv\int_{|1-v|}^{1+v}du\left(\frac{4v^2-(1-u^2+v^2)^2}{4uv}\right)^2 (uv)^{n_{\rm UV}} \overline{I^2(n_{\rm UV},k,v,u)}\,,
\end{align}
and has to be computed numerically. In order to have an idea of the magnitude of $F(n_{\rm UV})$ we provide a rough numerical fit for $n_{\rm UV}>0$, which reads
\begin{align}\label{eq:f_UV}
F(n_{\rm UV})\approx 11.5+3.25\frac{n_{\rm UV}^2}{\sqrt{16-n_{\rm UV}^2}}\,.
\end{align}

On the other hand, when $n_{\rm UV}>4$ the integral diverges for $v_p\to 0$. Thus, the relevant contribution comes from the lower cut-off at $v_p\ll1$. In this case, we find that
\begin{align}\label{eq:OGWUV2}
\Omega^{\rm UV}_{\rm GW}(k\gg k_{p},n_{\rm UV}>4)\approx \frac{4}{3}\left(\frac{1}{n_{\rm UV}-4}+\frac{1}{n_{\rm IR}+4}\right){\cal A}_{\cal R}^2\left(\frac{k}{k_p}\right)^{-4-n_{\rm UV}}\,,
\end{align}
where we included the contribution from the mixed regions for better accuracy.
The amplitude in the above approximations Eqs.~\eqref{eq:OGWUV} and \eqref{eq:OGWUV2} should be taken as order of magnitude estimates, specially when $n_{\rm UV}\sim 4$. For $n_{\rm UV}=4$ there is a logarithmic divergence that goes as $\ln({k}/{k_p})$. Most importantly though, they provide an accurate estimation of the power-law index of $k$. Let us note that while the spectral index of Eqs.~\eqref{eq:OGWUV} and \eqref{eq:OGWUV2} is already discussed in Ref.~\cite{Xu:2019bdp}, here we derive analytical formulas for the amplitude as well. 

We find that the induced GW spectrum for the broken power-law curvature perturbation \eqref{eq:PR} follows another broken power-law given by
\begin{align}\label{eq:OMGWR}
\Omega_{\rm GWs}(k)\propto{\cal A}^2_{\cal R}\left\{
\begin{aligned}
&\left(\frac{k}{k_p}\right)^3&k\ll k_p\\
&\left(\frac{k}{k_p}\right)^{-\Delta}& k\gg k_p
\end{aligned}
\right.\,,
\qquad
\rm with
\qquad
\Delta=\left\{
\begin{aligned}
&2n_{\rm UV} & 0<n_{\rm UV}<4\\
&4+n_{\rm UV}& n_{\rm UV}>4
\end{aligned} 
\right.\,.
\end{align}
With the template \eqref{eq:OMGWR} we may infer the value of $\fnl$ by looking at the high frequency tail of the induced GW spectrum from Eq.~\eqref{eq:nuvfnl}. Note that such estimate includes the underlying assumption that the curvature power spectrum is generated in single-field inflationary models and that the GWs are induced during radiation domination. Nevertheless, the analytical formulas \eqref{eq:OGWUV} and \eqref{eq:OGWUV2} are valid for any broken power-law primordial curvature power spectrum in a radiation dominated universe. We derive the corrections to the spectral tilt for general cosmological backgrounds in the next subsection.  

\begin{figure}[tpb]
\centering
\includegraphics[width=0.48\linewidth]{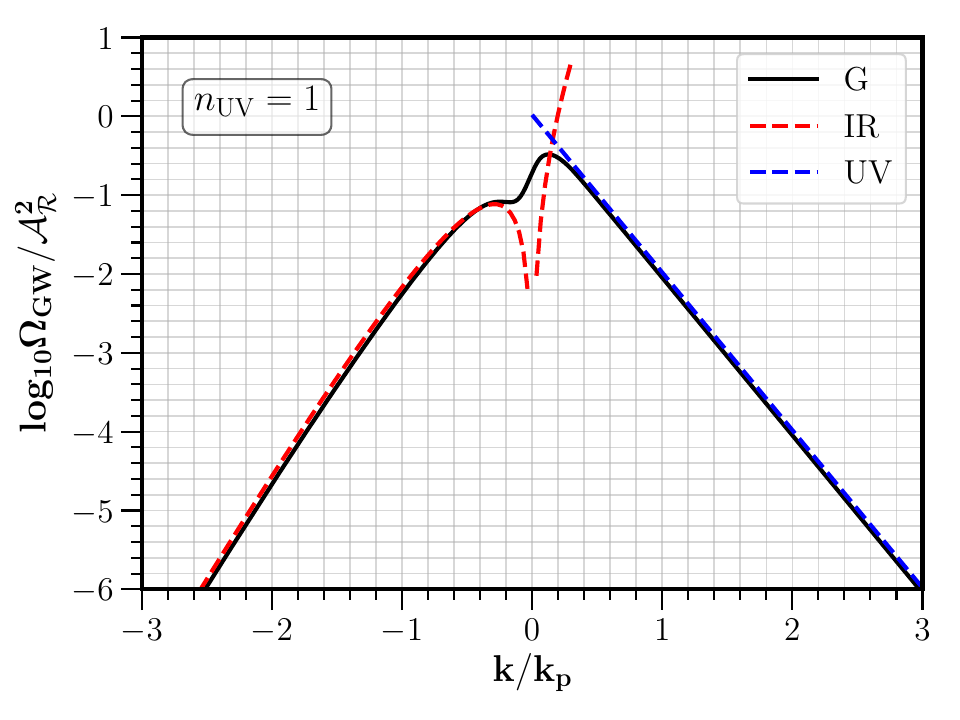}
\includegraphics[width=0.48\linewidth]{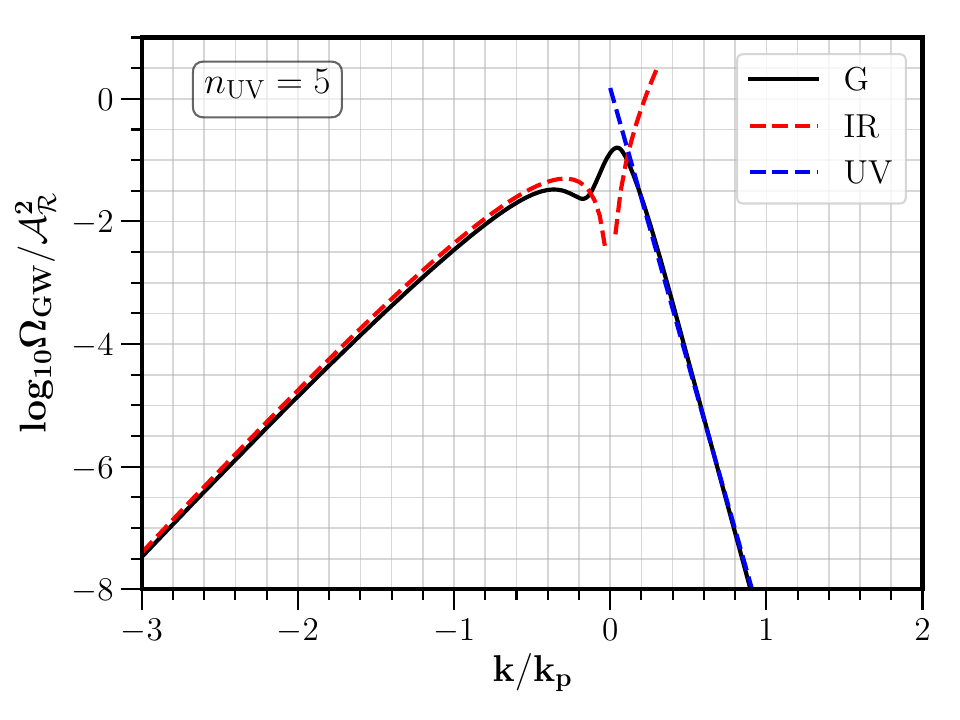}
\caption{ Induced GW spectrum from a broken power-law primordial spectrum \eqref{eq:PR} in terms of the wavenumber normalized to the peak position. In both figures we used $n_{\rm IR}=4$. The black line is the result of a numerical integration while the dashed lines correspond to the analytical approximations for the IR \eqref{eq:OGWIR} (in red) and the UV \eqref{eq:OGWUV} and \eqref{eq:OGWUV2} (in blue). On the left we consider the case $n_{\rm UV}=1$. On the right we plot the case $n_{\rm UV}=5$. See how the analytical approximations give a good estimate of the spectrum far from the peak. In both figures, the peak near $k\sim {\sqrt{3}}k_p/{2}$ is due to the divergence at $u+v=\sqrt{3}$ in the kernel \eqref{eq:kernelRD} in radiation domination. The physical origin of the divergence is a resonance that occurs the wavenumber $k$ of the tensor mode equals to two wavenumber $c_sk$ of the scalar modes. \label{fig:OMGW}}
\end{figure}

\subsection{Induced GWs in general cosmological backgrounds\label{subsec:generalcosmo}}

We shall generalize the previous template \eqref{eq:OMGWR} for the GW spectrum induced by a broken power-law to the case of general cosmological backgrounds. We assume that at the time of generation of the induced GWs, the universe was dominated by a perfect fluid with a constant equation of state parameter $w>-1/3$. In this case, the analytical approximations of the induced GW spectrum on the IR and UV limits is essentially analogous to the one explained in Sec.~\ref{subsec:GWpowerlaw}. The main differences are twofold. The first one lies in the $w$ dependence of the asymptotic behaviour of the kernel given in App.~\ref{app:kernelgeneral}. The second one is the presence of an additional scale $k_{\rm rh}$ which corresponds to the scale that last crossed the horizon at reheating. Taking into account the $w$ dependence and assuming an instantaneous reheating, we find that the induced GW spectrum is now two broken power-laws, concretely 
\begin{align}\label{eq:OMGWGB}
\Omega_{\rm GW}(k)\propto{\cal A}^2_{\cal R}\left\{
\begin{aligned}
&\left(\frac{k}{k_p}\right)^{3}&k\ll k_{\rm rh}\\
&\left(\frac{k}{k_p}\right)^{3-2|b|}&k_{\rm rh}\ll k\ll k_p\\
&\left(\frac{k}{k_p}\right)^{-\Delta-2b}& k\gg k_p
\end{aligned}
\right.\,,
\end{align}
where we introduced for convenience
\begin{align}
b=\frac{1-3w}{1+3w}\,.
\end{align}
When the universe is dominated by radiation, the equation of state is $w=1/3$ ($b=0$) and we recover Eq.~\eqref{eq:OMGWR}. It should be noted that the instantaneous reheating approximation only affects the transition around $k\sim k_{\rm rh}$. The conclusions on the spectral tilt for $k\gg k_{\rm rh}$ are unchanged by the instantaneous reheating. It is also important to note that the convergence conditions have changed due to the $w$ dependence. In particular, we now have that
\begin{align}
\Delta=\left\{
\begin{aligned}
&2n_{\rm UV} & 0<n_{\rm UV}<4\\
&4+n_{\rm UV}& n_{\rm UV}>4
\end{aligned} 
\right.\,.
\end{align}
The difference in the $w$ dependence of the IR and UV spectral tilts for $k\gg k_{\rm rh}$ in Eq.~\eqref{eq:OMGWGB} can be explained as follows. Modes with $k\gg k_{p}$ enter the horizon before the peak in the scalar spectrum. Then they evolve as usual and experience the redshift of the corresponding background \cite{Caprini:2018mtu}. This yields the factor $-2b$ in the exponent of the UV tail of Eq.~\eqref{eq:OMGWGB}. However, modes which enter after the peak, i.e. $k\ll k_{p}$, experienced a second superhorizon growth for $b<0$ ($w>1/3$) due to the fact that the source term of induced GWs decays slower than the background expansion \cite{Domenech:2020kqm}. This explains the factor $-2|b|$ in the exponent of the middle IR tail of Eq.~\eqref{eq:OMGWGB}.

The $w$ dependence of the UV tail of the GWs spectrum \eqref{eq:OMGWGB} may be a hindrance in disentangling the contribution coming from $\fnl$. The degeneracy between $\fnl$ and $b$ may be broken by the simultaneous observation of the IR tail of the induced GW spectrum. Despite the degeneracy of the IR spectral tilt for $|b|<1$ ($0<w<\infty$), by observing the IR tilt we may infer a value (or two) for the corresponding equation of state $w$. Using such estimation for $w$, we may extract a value (or two) of $\fnl$ from the UV tail.  

\section{Non-Gaussian corrections to the induced GWs spectrum \label{sec:nonGaussianities}}

In Sec.~\ref{sec:induced} we computed the GWs induced by a broken-power-law primordial curvature power spectrum given by Eq.~\eqref{eq:PR}. We used the fact that the spectral tilt of the UV tail of the primordial spectrum is related to the magnitude of the non-Gaussianity parameter $\fnl$ by Eq.~\eqref{eq:nuvfnl} to argue that $\fnl$ may be inferred from observations of the UV tail of the induced GW spectrum. The simple identification between $\fnl$ and the tilt of $\OmegaGW$ that we have described is simply due to the fact that the main features of the inflationary potential can be described by a single parameter. This means that with one measurement (in this case, of the linear perturbations) we can retrieve the important parts of the non-linear potential. There are however some purely non-Gaussian contributions to the induced GWs that we have not yet taken into account. These are due to the fact that the GWs are sourced by a 4-point correlation function, which can be grouped into a single or a set of contributions depending on the statistical nature of the perturbations. The impact of such non-Gaussianities has been previously estimated in Refs. \cite{Cai:2018dig,Unal:2018yaa,Ragavendra:2020sop,Yuan:2020iwf}. These works did not consider in generality an analytical relation between the amplitude of non-Gaussianities, the shape of the power spectrum and the abundance of PBHs, steps that we pursue in this section.

The intrinsically non-Gaussian effects on $\OmegaGW$ are twofold. The first one stands for a redefinition of the power spectrum \cite{Cai:2018dig,Yuan:2020iwf}, and corresponds at leading order in $\fnl$ to the so-called hybrid diagram in Ref.~\cite{Unal:2018yaa}. The contribution to the induced GWs is analogous to the Gaussian calculation of Sec.~\ref{sec:induced} but with a modified primordial spectrum given by
\begin{align}\label{eq:PRNG}
{{\cal P}_{\cal R}(k)}={{\cal P}_{{\cal R}_G}(k)}+F_{\rm NL}^2\int_0^\infty d v\int_{|1- v|}^{1+ v}\frac{d u}{ v^2  u^2} {{\cal P}_{{\cal R}_G}(k  v)}{{\cal P}_{{\cal R}_G}(k u)}\equiv{{\cal P}_{{\cal R}_G}(k)}+{{\cal P}^{NG}_{{\cal R}_G}(k)}\,.
\end{align}
where we have also defined
\be
F_{\rm NL}=\frac{3}{5}\fnl
\ee
in order to avoid carrying (5/3) factors. This contribution stands for only one of the contractions of the 4-point function. Indeed, the calculation of Sec.~\ref{sec:induced} assumes that the connected 4-point correlation function is vanishing and that one can evolve two pairs of scalar modes, each pair inducing GWs, independently. In this way the contribution that we have considered stands as a modification of the primordial curvature power spectrum. However, in general the connected 4-point function is non-vanishing. At leading order in $\fnl$, the connected contribution can be represented by a walnut diagram as in Ref.~\cite{Unal:2018yaa}. A non-vanishing connected 4-point correlation function essentially takes into account the correlation of two pairs of scalar modes during their evolution.

As we proceed to show, the inclusion of the hybrid and walnut diagrams has little impact in the IR and UV tails of the GW spectrum induced by the broken power-law curvature spectrum \eqref{eq:PR}. In App.~\ref{app:calculations} we present the general formulas for the Gaussian, the hybrid and the walnut contributions to the induced GW spectrum. To simplify the discussion we shall focus solely on the modification of the primordial curvature power spectrum or the hybrid diagram. We present the analytical formulas for the walnut contribution in the IR and UV limits for a broken power-law primordial spectrum in App.~\ref{app:walnut}. We find that the walnut contribution is negligible in the IR limit but may dominate over the hybrid in the UV limit. Nevertheless, this does not affect the spectral tilt of the UV limit. We also note that the walnut diagram is exactly vanishing after integration due to symmetries in the case of ${\cal P}_{{\cal R}_G}(k)\propto k^3$ and ${\cal P}_{{\cal R}_G}(k)\propto k^5$.

\begin{figure}[tpb]
\centering
\includegraphics[width=0.48\linewidth]{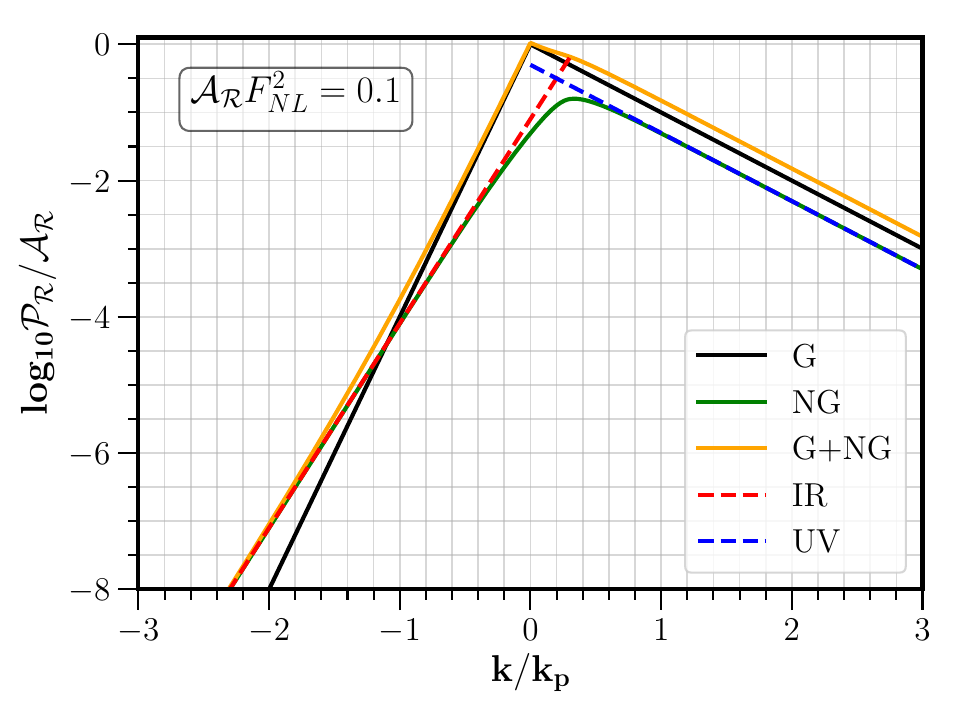}
\includegraphics[width=0.48\linewidth]{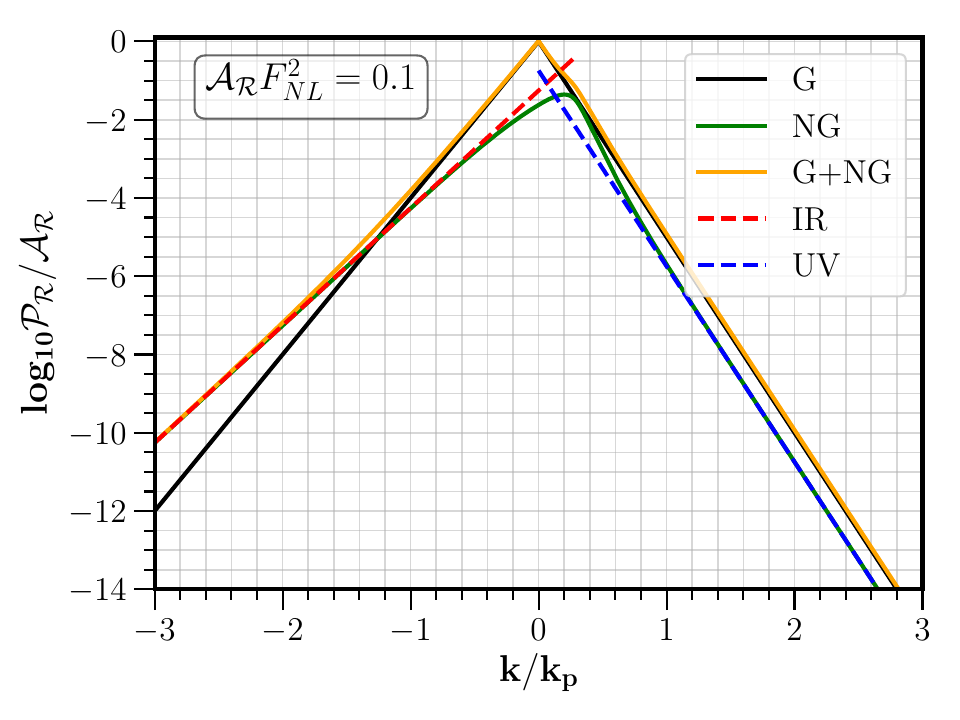}
\caption{Non-Gaussian corrections to the power spectrum generated in single-field inflation for different steepness of the decay: respectively for $
n_{\rm UV}=1$ and $n_{\rm UV}=5$ in the left and right panel. In both figures we used $n_{\rm IR}=4$ and ${\cal A}_{\cal R}\fnlsq=0.1$. In solid black and green we show the Gaussian and the non-Gaussian contribution to the spectrum \eqref{eq:PR} and \eqref{eq:PRNG}. The solid orange line is the sum of the two contributions. Dashed lines correspond to the analytical approximations \eqref{eq:NGcont} and \eqref{eq:NGcontuv} respectively for the IR (in red) and the UV (in blue). See how the analytical approximations capture the main essence of the IR and UV tails.  \label{fig:NG_PS}}
\end{figure}

Let us show that the modification of the primordial spectrum due to non-Gaussianities is negligible for practical purposes in the broken power-law case. We analytically integrate Eq.~\eqref{eq:PRNG} in the IR and UV limits using the same procedure as explained in Sec.~\ref{subsec:GWpowerlaw}. On one hand, we find that in the IR limit (with $n_{\rm IR}>3/2$) the non-Gaussian contribution in Eq.~\eqref{eq:PRNG} is given by
\begin{align}\label{eq:NGcont}
{{\cal P}^{NG}_{{\cal R}_G}(k\ll k_p)}\approx2\left(\frac{1}{2n_{\rm IR}-3}+\frac{1}{2n_{\rm UV}+3}\right)F_{\rm NL}^2{\cal A}_{\cal R}^2\left(\frac{k}{k_p}\right)^3\,.
\end{align}
It is interesting to note that at some point the amplitude of the IR tail of the initial spectrum is dominated by the non-Gaussian contribution \eqref{eq:NGcont} as it decays slower than the Gaussian contribution. This occurs roughly at
\begin{align}
\frac{k_{\rm NG}}{k_p}\approx\left({2{\cal A}_{\cal R}F_{\rm NL}^2}\left(\frac{1}{2n_{\rm IR}-3}+\frac{1}{2n_{\rm UV}+3}\right)\right)^{\frac{1}{n_{\rm IR}-3}}\ll1\,,
\end{align}
where in the last step we used that ${\cal A}_{\cal R}F_{\rm NL}^2\ll1$. Nevertheless, even though the non-Gaussian contribution eventually dominates, it does not change the IR spectral tilt of the GW spectrum as it can be seen from Eq.~\eqref{eq:OGWIR} replacing $n_{\rm IR}$ by $3$. On the other hand, we find that the UV tail is well approximated by 
\begin{align}\label{eq:NGcontuv}
{{\cal P}^{NG}_{{\cal R}_G}(k\gg k_p)}\approx4\left(\frac{1}{n_{\rm UV}}+\frac{1}{n_{\rm IR}}\right){F_{\rm NL}^2{\cal A}_{\cal R}^2}\left(\frac{k}{k_p}\right)^{-n_{\rm UV}}\,,
\end{align}
where for an exact flat spectrum, that is $n_{\rm UV}=0$, one has a logarithmic dependence in $k/k_p$. We conclude that in the UV regime non-Gaussianity slightly increases the amplitude of the resulting spectrum. 

We show the non-Gaussian correction to the power spectrum in Fig.~\ref{fig:NG_PS}. Note that the expansion parameter is ${\cal A}_{\cal R}F_{\rm NL}^2$, and thus we require ${\cal A}_{\cal R}F_{\rm NL}^2\ll1$. In Fig.~\ref{fig:NG_PS} we fixed ${\cal A}_{\cal R}F_{\rm NL}^2$ = 0.1 so that the effects of the non-Gaussian contribution are more evident. As is clear from Fig.~\ref{fig:NG_PS} the UV tail of the primordial spectrum is barely affected by the non-Gaussian contribution. Moreover, as we argue in the next section,  ${\cal A}_{\cal R}F_{\rm NL}^2\ll1$ is always satisfied provided that there is no overproduction of PBHs, even when $F_{\rm NL}\gg1$. Thus, the fact that ${\cal A}_{\cal R}F_{\rm NL}^2\ll1$ together with Eqs.~\eqref{eq:NGcont} and \eqref{eq:NGcontuv} showing that the spectral features in the IR and UV are maintained, justify why the non-Gaussian contributions can be neglected (as we did in Sec.~\ref{sec:induced}).

\subsection{Can there be PBHs in the non-perturbative regime?}

We might ask whether the perturbative constraint  ${\cal A}_{\cal R}F^2_{NL}\ll1$ holds in the regime of a large number of PBHs (large ${\cal A}_{\cal R}$) and/or large $F_{\rm NL}$. As we will see, perturbativity is always satisfied provided that PBHs are not overproduced. If the statistics of peaks are Gaussian, then the number of PBHs at formation is roughly
\be
\beta_{\rm G} \simeq e^{-{\cal R}_c^2/(2\sigma^2)} \ ,
\ee
where ${\cal R}_c$ is the critical value for an overdensity to collapse into a BH and $\sigma$ is the variance of the perturbation. PBHs are not overproduced  if $\nu\equiv {\cal R}_c/\sigma \gtrsim 6$. Since ${\cal R}_c \sim 0.5$, we need  $\sigma^2 \sim {\cal A}_{\cal R} \lesssim \mathcal{O}(0.01)$. In turn,  this implies that there is a potential problem for $F_{\rm NL} \gtrsim 10$ if ${\cal A}_{\cal R}\sim 0.01$ , since then ${\cal A}_{\cal R} F^2_{\rm NL} \sim 1$, and the expansion that we have used might not longer be valid.

This is however never the case, if we take into account how the threshold for collapse varies with $F_{\rm NL}$. In non-Gaussian theories described by a local function of a Gaussian variable, ${\cal R}=F({\cal R}_G)$, the abundance of PBHs can also be written in terms of the Gaussian underlying field. If we denote $\mu_*$ the amplitude of the Gaussian field such that the non-Gaussian overdensity is at the threshold of collapse, i.e, ${\cal R}_c=F(\mu_*)$, then the abundance of PBHs at formation is given by
\be\label{eq:beta_NG}
\beta_{\rm NG} \simeq e^{-\mu_*^2/(2\sigma^2)}  \ .
\ee
For the case of single-field models of inflation, it has been found that, for $F_{\rm NL} > 2$, $\mu_* \simeq 1/ (2 F_{\rm NL})$ \cite{Atal:2019erb}. This means that, for a given constant abundance $\beta_{\rm NG}\sim e^{-\nu_*^2}$, we have that
\be
{\cal A}_{\cal R} F^2_{\rm NL} \sim \frac{1}{4\nu_*^2} \ll 1 \ . 
\ee 
This implies that the effect of loops induced by non-Gaussianities are hardly directly measurable,  and that large regions of the parameter space studied in Refs. \cite{Cai:2018dig,Unal:2018yaa,Yuan:2020iwf} would not be consistent with an inflationary origin of the perturbations.

Note that in this justification for using the perturbative template of non-Gaussianities for the computation of the induced GWs we have used the \emph{non-perturbative} information that $\mu_* \simeq 1/(2 F_{\rm NL})$ for large $F_{\rm NL}$. Had we used the perturbative quadratic template for calculating ${\cal R}_c$ at large $F_{\rm NL}$ we would had find a different scaling \cite{Atal:2019erb}. We also note that for Eq. (\ref{eq:beta_NG}) to hold, we need to require that maxima of the Gaussian field are identified with maxima of the non-Gaussian field. This is indeed the case for the logarithmic template of Eq. (\ref{eq:ng_np}), but not for its quadratic truncation Eq. (\ref{eq:ng_pert}), as used e.g. in \cite{Riccardi:2021rlf}\footnote{This identification also holds for maxima of the curvature and density perturbations if the power spectrum is peaked \cite{Yoo:2018kvb}.}. In general, quantities depending on extreme realizations of the perturbations, as the shape of rare peaks and their abundance, are more sensitive to the non-perturbative completion of the non-Gaussianities, while for quantities depending on the bulk of the perturbations, as the induced GWs, a perturbative description might suffice. 

\section{Observational Considerations \label{sec:observations}}

We now discuss the prospects for detecting $\OmegaGW$ in this setup. In the following we constrain the amplitude and position of the peak of the power spectrum as a function of the decay of the power spectrum (or equivalently $\fnl$). 

\subsection{NANOGrav as the UV tail of \texorpdfstring{$\OmegaGW$}{omegaGW}}

The NANOGrav collaboration recently reported the presence of signal consistent with a stochastic background of gravitational waves \cite{Arzoumanian:2020vkk}. This signal can correspond to the peak of induced GWs, as shown by several authors \cite{Vaskonen:2020lbd,DeLuca:2020agl,Kohri:2020qqd,Bian:2020bps,Sugiyama:2020roc,Domenech:2020ers,Bhattacharya:2020lhc,Inomata:2020xad}\footnote{The signal could also correspond to the background resulting from the past mergers of large or 'stupendously' large black holes \cite{Arzoumanian:2020vkk,Middleton:2020asl,Atal:2020yic}.}. Another possibility is that the signal comes from the UV tail of induced GWs. If this is the case, then the observational bounds on the spectral index bounds the spectral index of the power spectrum and thus also the amplitude of $\fnl$. Indeed, if the signal is modelled as a power-law, i.e.
\be
\OmegaGW(k)=A_p\left(\frac{k}{k_{p}}\right)^{-\xi} \ ,
\ee
then 
\be\label{eq:xi_constraints}
\xi\in(-0.5,1.5)
\ee
at 1-$\sigma$ confidence level. Then using Eq.~\eqref{eq:nuvfnl} this implies that 
\be
\fnl<0.3 \ .
\ee
This is a very strong constraint of $\fnl$, and shows that the non-Gaussianities in this case are negligible. Note, however, that this bound is sensitive to the value of the equation of state of the universe $w$. For $w\neq1/3$ we have instead 
\begin{align}
-0.1<\fnl+\frac{5}{12}\frac{1-3w}{1+3w}<0.3\,.
\end{align}
This implies that at most we could have $\fnl\lesssim 0.7$, if we consider the unrealistic limit $w\to\infty$. Thus, we conclude that even taking into account the possible degeneracy with the equation of state parameter $w$, that is assuming we cannot see the IR tail to break the degeneracy, NANOgrav results imply a low value for $\fnl$. 

In passing, let us comment the differences with Ref.~\cite{Domenech:2020ers} which also considers GWs induced by a sharply peaked primordial spectrum and a general equation of state as an explanation for the NANOGrav results. Contrary to this work, Ref.~\cite{Domenech:2020ers} uses the middle IR tilt of Eq.~\eqref{eq:OMGWGB} to fit the NANOGrav results. This yields two main differences. First, the dependence on the equation of state is different by the absolute value of $\beta$. However, in the exact limit $\fnl=0$ we recover the bound on $w<1/3$ from Ref.~\cite{Domenech:2020ers}. Second, for $w<0$, the absolute peak of the induced GW spectrum is at the reheating scale $k_{\rm rh}$, although for a perfect fluid there is another peak at around $k\sim 2c_sk_p$ where $c_s$ is the speed of scalar fluctuations. In other words, if we fix the absolute peak of the GW spectrum at $k\sim k_{\rm rh}$ to fit the NANOGrav results at $f\sim 10^{-9}\,{\rm Hz}$, then the peak in the primordial spectrum is actually at much smaller scales than the peak of the induced GW spectrum since $k_p\gg k_{\rm rh}$. This implies that the corresponding PBHs formed from the collapse of primordial fluctuations have a mass much smaller than solar mass. Instead in this work, to fit the NANOGrav results with the UV tail of the GW spectrum, we fix the peak of the primordial spectrum to be at or below to the scale corresponding to a frequency of $10^{-9}{\rm Hz}$. Thus, the resulting PBHs have a mass of tenth of solar masses or larger.

\begin{figure}[tpb]
\centering
\includegraphics[width=0.7\linewidth]{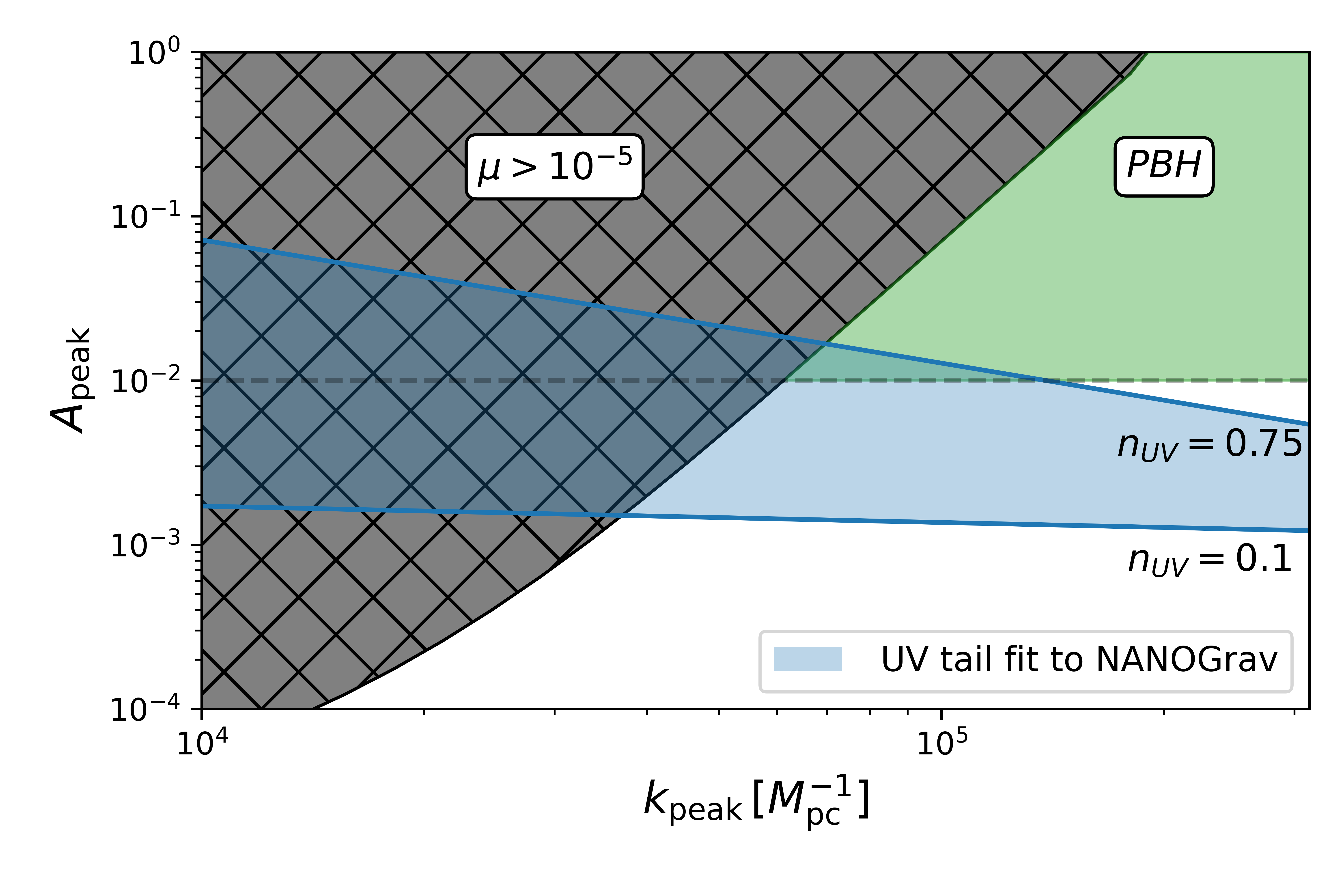}
\caption{Constraints for $A_p$ and $k_p$, for models having a slope of the UV tail consistent with NANOgrav. In grey we show the region ruled out by $\mu$ distortion constraints and in green the region where sizeable amount of PBHs are produced (that holds whenever $\fnl$ is not extremely large, as in the present case).} \label{fig:nano}
\end{figure}

Since the peak of the power spectrum is at lower frequencies than those probed by NANOGrav, then there are further constraints coming from $\mu$ distortions and from overproduction of PBHs. The $\mu$-distortions are also sensitive to the shape of the power spectrum. They take the form \cite{Chluba:2015bqa}
\be
\mu\simeq 2.3 \int_{k_0=1}^{\infty} d \log k\, \Pk W(k)  \ ,
\ee
where $k$ is units of ${\rm Mpc}^{-1}$, and the window function $W(k)$ is given by
\be
W(k)=\left[\exp\left(-\frac{\left[\frac{k}{1360}\right]^2}{1+\left[\frac{k}{260}\right]^{0.3}+\frac{k}{340}}\right)-\exp\left(-\left[\frac{k}{32}\right]^{2}\right) \right] \ .
\ee
In Figure \ref{fig:nano} we show the constraints for $A_p$ and $k_p$, for models having a slope of the UV tail consistent with \eqref{eq:xi_constraints}. We show also the region that is ruled out by $\mu$ distortions constraints \cite{Fixsen:1996nj} and the region resulting, for small non-Gaussianities, in possibly large fraction of PBHs ($A_p>0.01$).
This interplay between gravitational wave and spectral distortions has been considered in \cite{Kite:2020uix}. 

\subsection{Present and future constraints}

\begin{figure}[tpb]
\centering
\includegraphics[width=0.48\linewidth]{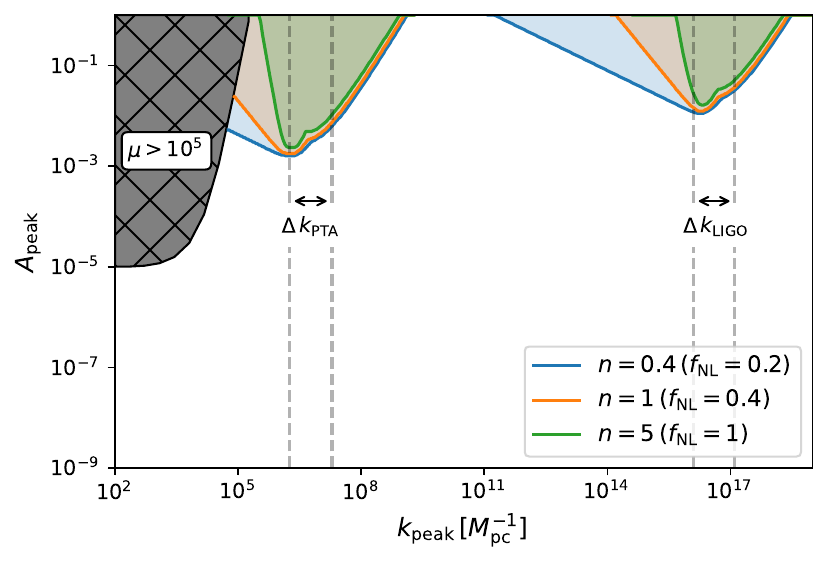}
\includegraphics[width=0.48\linewidth]{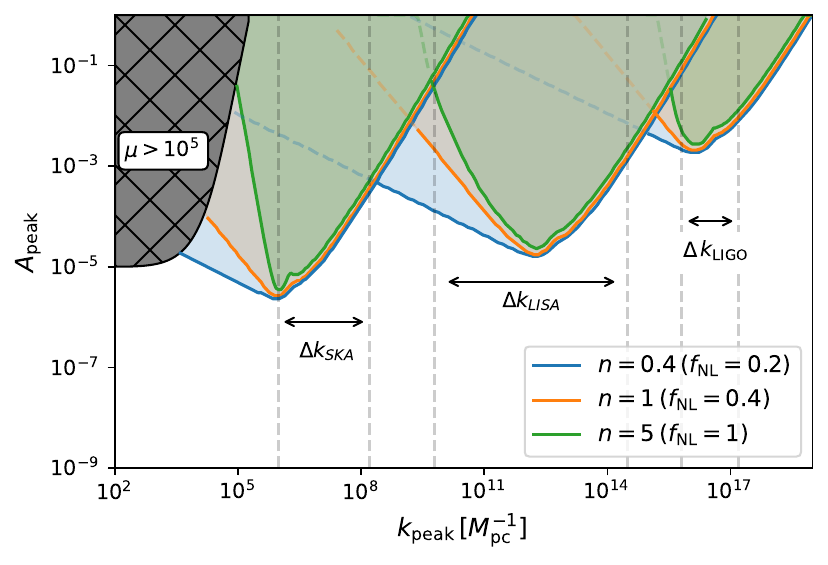}

\caption{Constraints for $A_p$ and $k_p$ coming from present (left pannel) and future (right pannel) experiments, as well as $\mu$ distortions constraints. For small decay of the power spectrum ($n_{\rm UV}$) or equivalently small $\fnl$, observations have a larger constraining power.}. \label{fig:further_exp}
\end{figure}

It is more likely to detect a flat GW spectrum than a spiky one, since the amplitude of the tail for a flat spectrum is larger for a larger number of decades in k-space. Then, in single-field inflation, it is more likely to detect models with low $\fnl$. 
In Figure \ref{fig:further_exp} we show how the constraints on GW coming from PTA and LIGO 02 run put bounds on the amplitude of power spectrum as a function of its decay of the power spectrum $n_{\rm UV}$, or, equivalently, $\fnl$. We also show how these constraints will improve with future experiments such as SKA \cite{Janssen:2014dka} and LISA \cite{Audley:2017drz}, as well as with the design sensitivity of LIGO.

If the power spectrum is flat enough after the peak, then the constraining power of these experiments is enhanced. Actually, for sufficiently flat power spectra, $n_{\rm UV}<5$ or $\fnl<2$, a given power spectra can be detected by the two experiments. This coincident regime corresponds to the region above the dotted lines in Figure \ref{fig:further_exp}.
 
Let us point out that these constraints should be taken with caution since we might expect changes in the curvature of the potential some decades after the peak. If these contributions make the potential flatter (as might be the case if the bump is sufficiently localized), then these constraints are a lower bound. However, if the potential is steeper as we go away from the peak (as might happen if the peak is rather flat), then these are upper bounds.\\

Finally, let us note that for $n_{\rm UV}>5$ or equivalently $\fnl>2$, we recover the constraints coming from a power spectrum with a hard cut-off after the peak, as has been considered e.g. in Refs. \cite{Byrnes:2018txb,Inomata:2018epa}. 

\section{Conclusions and Discussion \label{sec:conclusions}}

In this work, we have showed how the induced GW spectrum probes the level of non-Gaussianity of primordial curvature fluctuations, if these are generated during single-field inflation. To have a detectable induced GW signal, the primordial curvature power spectrum on small scales must be enhanced with respect to the one measured in CMB scales. If such enhancement occurs by a bump in the single-field potential, the primordial power spectrum takes the shape of a broken power-law. Rather generally, the spectral index of the UV power-law is directly related to the non-Gaussianity parameter $\fnl$. Such broken power-law in the primordial power spectrum, translates into a broken power-law for the induced GW spectrum.
While the spectral tilt of the induced GWs is universal at IR scales, that is for $k\ll k_p$ where $k_p$ corresponds to the scale of the peak in the primordial spectrum, the spectral tilt at UV scales ($k\gg k_p$) is directly linked to $\fnl$ by
\begin{align}\label{eq:PR_concl}
\Omega_{\rm GWs}(k\gg k_p)\propto{\cal A}^2_{\cal R}\left(\frac{k}{k_p}\right)^{-\Delta} \,,
\qquad
\rm with
\qquad
\Delta=\left\{
\begin{aligned}
&2n_{\rm UV} & 0<n_{\rm UV}<4\\
&4+n_{\rm UV}& n_{\rm UV}>4
\end{aligned} 
\right.\,,
\end{align}
where
\be
n_{\rm UV}=\frac{12}{5} \fnl \,,
\ee
is the spectral tilt of the UV power-law of the primordial power spectrum. In this way, by measuring the UV tail of the induced GW spectrum, we can directly measure $\fnl$. The proportionality constant of Eq.~\eqref{eq:PR_concl} can be found in Eqs.~\eqref{eq:f_UV}-\eqref{eq:OGWUV2}. We also studied the generalization of Eq.~\eqref{eq:PR_concl} to general cosmological backgrounds, given in Eq.~\eqref{eq:OMGWGB}. We found that if GW are induced during an epoch where the universe is not dominated by radiation, the IR and UV spectral tilts of the induced GW spectrum are modified. However, the degeneracy between $\fnl$ and the equation of state parameter $w$ is broken by the simultaneous observation of the IR and UV slopes of the GW spectrum. 

Such link between the UV spectral tilt and the level of non-Gaussianity around the scale of the peak or smaller, led us to study the impact of such non-Gaussianity in the induced GW spectrum. We found that in general the presence of primordial non-Gaussianities does not alter the spectral tilt of the induced GW spectrum. The main effects of the non-Gaussianity are twofold: $(i)$ a modification of the primordial spectrum given by Eq.~\eqref{eq:PRNG} as studied in Ref.~\cite{Cai:2018dig} and $(ii)$ a correlation between two pairs of scalar modes during their evolution. The latter corresponds to the connected $4$-point function and was pointed out in Ref.~\cite{Unal:2018yaa}. We showed in Sec.~\ref{sec:nonGaussianities} and in App.~\ref{app:walnut} that, in the case of a broken power-law, the only effect of non-Gaussianities is a small broadening of the peak of the induced GW spectrum. Nevertheless, the broadening of the peak is negligible provided there is no overproduction of PBHs. Thus, our results show that the spectral features, including the tilt of the UV slope, is robust to non-linearities in the inflationary sector. Although we only focused on the local type non-Gaussianity of the bispectrum, justified in this setup by the perturbativity conditions holding from non-overproduction of PBHs, we expect that a local type non-Gaussianity of the trispectrum, often referred to as $g_{\rm NL}$, would not change our results. We leave this issue for future work.

It is of course important to determine whether the spectral features of single-field inflation are unique or not. Large amplifications of the power spectrum leading to observable induced GWs can also be obtained in multifield scenarios \cite{Clesse:2015wea,Palma:2020ejf,Fumagalli:2020adf, Braglia:2020eai,Braglia:2020taf,Fumagalli:2020nvq}. A measurement of the UV tail of the GW spectrum would definitively need a more detailed analysis in order to discriminate between these models. However it is conceivable that by having access to the detailed shape around the peak one may be able to distinguish them. The simplest scenario to discriminate with is if the amplification in the multifield case results from a rapidly varying turning rate in the inflaton trajectory. In this case there are several oscillations in the peak, as shown in Refs.~\cite{Braglia:2020taf,Fumagalli:2020nvq}, which are clearly distinguishable from the single-field case discussed here. On the other hand, if the turn rate is smooth enough, then the GWs has a wide and smooth peak, as in Ref.~\cite{Braglia:2020eai}. In contrast, for a power-law primordial spectrum the peak in the induced GW spectrum has two hills, as is clear from Fig.~\ref{fig:OMGW}, unless $n_{\rm UV}<1$. 

It is also interesting to consider the differences with the induced GWs from possible resonances during and after inflation. If the curvature power spectrum is enhanced due to a narrow resonance \cite{Chen:2019zza}, that leads to a very sharp, then the IR tail of the induced GW spectrum decays as $k^2$. If more very sharp peaks are present in the primordial curvature power spectrum then the induced GW spectrum also present a series of spikes near the peak \cite{Cai:2019amo}. This characteristic signatures distinguish these models from the one studied in this work. Nevertheless, a close degeneracy might occur for a resonance during inflation as in Ref.~\cite{Zhou:2020kkf}, since the resulting induced GW spectrum also resembles a broken power-law. However, in the case of Ref.~\cite{Zhou:2020kkf} the generation occurs during inflation and therefore lacks the characteristic peak of radiation domination at $k/k_{p}\sim 2/\sqrt{3}$. This fact and the possible impact of non-gaussianities probably breaks the degeneracy. Also note that we have not considered any possible resonances after inflation due to a time dependent speed of GWs \cite{Cai:2020ovp}. We leave further investigation of these issues for future work.

Another interesting aspect is the partial degeneracy existing between the UV slopes of the induced GWs in single-field inflation
and the UV slope of GW backgrounds from other sources (see Ref.~\cite{Kuroyanagi:2018csn} for a compendium). The largest UV tilt reported in Ref.~\cite{Kuroyanagi:2018csn} corresponds to GWs created by sound waves in the primordial plasma \cite{Hindmarsh:2017gnf}, for which $\Delta=4$, corresponding in the single-field case to $\fnl=0.8$. Models featuring a power-law in the UV are thus degenerate with rather mild non-Gaussianities\footnote{On the other hand, a very sharp cut-off in the UV can also be realised, as in the UV tail of GW from black hole binaries  \cite{Shannon:2015ect}.}, $\fnl \lesssim 0.8$, corresponding to $\epsilon_2^{ge}<2$ or $|\eta_V|<16$. Flat spectra can also be realised, as in scenarios with cosmic (super)strings. A small tilt $\Delta \in (0,0.2)$ could be obtained at intermediate regions, as in the examples considered in Ref.~\cite{Blanco-Pillado:2021ygr},  which are thus degenerate with $\fnl \in(0,0.04)$  in the single-field inflationary case.  Note that a modified thermal history of the Universe also modifies the UV spectral index in these cases \cite{Cui:2018rwi,Ramberg:2019dgi,Auclair:2019wcv}. In fact, the additional exponent $-2b$ that we encounter in Eq. \eqref{eq:OMGWGB} would also be present in these scenarios, as this stands for changes in the expansion history  that affects the propagation of GWs irrespective of their origin.

While a single measurement of $\Delta$ in these coincident regions would not allow to disentangle between these cases, if, for a given thermal history, $\Delta$ lies outside of it then this would be a strong case for the inflationary origin, as suggested in Ref.~\cite{Liu:2020oqe}. Such degeneracies are broken by the additional observation of the IR tail and/or the peak of the GW spectrum and, of course, by the detection of the possibly sizable PBH counterpart to the induced GWs.

In the last section, by virtue of the relation between the tilt  and $\fnl$, we have made a few considerations on present and future experiments. For example, if NANOGrav corresponds the UV tail of induced GWs, then $\fnl<0.3$ for a radiation dominated Universe. Interestingly, we found that independent of the equation of state parameter $w$, the upper bound on $\fnl$ is rather small with $\fnl<0.7$. We have also commented on future SKA and LISA measurements and show how the detectability of single-field inflationary models largely increments if $\fnl$ is small.

\section*{Acknowledgements}
We thank Yann Gouttenoire and Caner Ünal for spotting a typo in the kernel function. We thank Jose J. Blanco-Pillado, Jaume Garriga, Shi Pi and Misao Sasaki for discussions. We also thank Matteo Braglia for useful comments. V.A. is supported in part by the Spanish Ministry MCIU/AEI/FEDER grant (PGC2018-094626-BC21) and the Basque Government grant (IT-979-16). G.D. as a Fellini fellow is supported by the European Union’s Horizon 2020 research and innovation programme under the Marie Sk{\l}odowska-Curie grant agreement No 754496. V.A. would like to thank Ferruccio Poli from Universita di Bologna for technical support.

\appendix

\section{Details on the calculations\label{app:calculations}}

In this appendix we provide the basic formulas for the calculations of induced GWs including the corrections due to primordial non-Gaussianity.
The calculation starts with the equations of motion for tensor modes at second order in cosmological perturbation theory, which in conformal time and in Fourier modes reads
\begin{align}\label{eq:forh}
h_{k,\lambda}''+2{\cal H}h_{k,\lambda}+k^2h_{k,\lambda}=S_\lambda(\mathbf{k})\,.
\end{align}
If we focus solely on the scalar squared terms of the source, we find that
\begin{align}
S_\lambda(\mathbf{k})=8\int \frac{d^3q}{(2\pi)^3}e_\lambda^{ij}(k)q_iq_j\Phi_q\Phi_{|\mathbf{k}-\mathbf{q}|}f(k\tau,q,|k-q|)\,,
\end{align}
where $\Phi_q$ refers to the primordial value of $\Phi$ by
\begin{align}
\Phi(q,\tau)=\Phi_q\,T(q\tau)\,,
\end{align}
where $T(q\tau)$ is the transfer function given by the first order equations of motion for $\Phi$ and
\begin{align}
f&(\tau,q,|\mathbf{k}-\mathbf{q}|)\nonumber\\& \equiv T(q\tau)T(|\mathbf{k}-\mathbf{q}|\tau)+\frac{2}{3\left(1+w\right)}\left[T(q\tau)+\frac{T'(q\tau)}{\cal H}\right]\left[T(|\mathbf{k}-\mathbf{q}|\tau)+\frac{T'(|\mathbf{k}-\mathbf{q}|\tau)}{\cal H}\right]\,.
\end{align}

Using the Green's function method one may find formal solutions to Eq.~\eqref{eq:forh}. With such solution, the two point function of the tensor modes is readily given by
\begin{align}
\langle h_\lambda(k,\tau)h_\lambda(k',\tau)\rangle=\int_0^\tau d\tau_1\int_0^\tau d\tau_2\frac{1}{kk'}G(k\tau,k\tau_1)G(k'\tau,k'\tau_2)\langle S_\lambda(k,\tau_1)S_\lambda(k',\tau_2)\rangle\,.
\end{align}
Note that here we have the expectation value of the source term squared which is related to the four point function of $\Phi$, namely
\begin{align}\label{eq:sl}
\langle S_\lambda(k,\tau_1)S_\lambda(k',\tau_2)\rangle=8^2\int \frac{d^3q}{(2\pi)^3}\int \frac{d^3q'}{(2\pi)^3}&e_\lambda^{ij}(k)q_iq_je_\lambda^{ij}(k')q'_iq'_jf(\tau_1,q,|\mathbf{k}-\mathbf{q}|)f(\tau_2,q',|\mathbf{k}'-\mathbf{q}'|)\nonumber\\&
\times\langle\Phi_q\Phi_{|\mathbf{k}-\mathbf{q}|}\Phi_{q'}\Phi_{|\mathbf{k}'-\mathbf{q}'|}\rangle\,.
\end{align}

Let us now consider that $\Phi$ has a small primordial non-Gaussianity component of the local form, which is usually parametrized by
\begin{align}\label{eq:FNLPHIExpansion}
\Phi_q=\Phi_q^g+\tilde \fnl\int \frac{d^3l}{(2\pi)^3}\Phi_l^g\Phi_{|\mathbf{q}-\mathbf{l}|}^g\,.
\end{align}
Using such expansion, we find that up to leading order in $\fnl$ the two point function of the source term is given by
\begin{align}\label{eq:sl2}
&\langle S_\lambda(k,\tau_1)S_\lambda(k',\tau_2)\rangle=8^2\int \frac{d^3q}{(2\pi)^3}\int \frac{d^3q'}{(2\pi)^3}e_\lambda^{ij}(k)q_iq_je_\lambda^{ij}(k')q'_iq'_jf(\tau_1,q,|\mathbf{k}-\mathbf{q}|)f(\tau_2,q',|\mathbf{k}'-\mathbf{q}'|)\nonumber\\&
\times\Bigg\{\langle\Phi^g_q\Phi^g_{|\mathbf{k}-\mathbf{q}|}\Phi^g_{q'}\Phi^g_{|\mathbf{k}'-\mathbf{q}'|}\rangle+\tilde \fnlsq\Bigg(\int\frac{d^3l}{(2\pi)^3}\int\frac{d^3l'}{(2\pi)^3}\langle\Phi^g_{q}\Phi^g_l\Phi^g_{|\mathbf{k}-\mathbf{q}-\mathbf{l}|}\Phi^g_{q'}\Phi^g_{l'}\Phi^g_{|\mathbf{k}'-\mathbf{q}'-\mathbf{l}'|}\rangle\nonumber\\&+(|\mathbf{k}-\mathbf{q}|\leftrightarrow q)+(|\mathbf{k'}-\mathbf{q'}|\leftrightarrow q')+(q\leftrightarrow q';|\mathbf{k}-\mathbf{q}|\leftrightarrow|\mathbf{k'}-\mathbf{q'}|)\Bigg)\Bigg\}\,.
\end{align}
We shall split Eq.~\eqref{eq:sl} into three contributions following the notation of Ref.~\cite{Unal:2018yaa}: the Gaussian, the hybrid and the walnut. The walnut diagram corresponds to the leading order of the connected 4-point function of $\Phi_q$. After Wick contractions and some algebra we arrive at
\begin{align}
&\langle S_\lambda(k,\tau_1)S_\lambda(k',\tau_2)\rangle=(2\pi)^3\delta(\mathbf{k}+\mathbf{k'})\nonumber\\&\times\left\{\langle S_\lambda(k,\tau_1)S_\lambda(k',\tau_2)\rangle'_g+\langle S_\lambda(k,\tau_1)S_\lambda(k',\tau_2)\rangle'_{NG,h}+\langle S_\lambda(k,\tau_1)S_\lambda(k',\tau_2)\rangle'_{NG,w}\right\}\,,
\end{align}
where
\begin{align}
\langle S_\lambda(k,\tau_1)&S_\lambda(k',\tau_2)\rangle'_g\equiv{8^2(2\pi^2)}\int \frac{d^3q}{2\pi} \left(e_\lambda^{ij}(k)q_iq_j\right)^2\nonumber\\&\times f(\tau_1,q,|\mathbf{k}-\mathbf{q}|)f(\tau_2,q,|\mathbf{k}-\mathbf{q}|)\frac{{\cal P}_\Phi(q)}{q^3}\frac{{\cal P}_\Phi(|\mathbf{k}-\mathbf{q}|)}{|\mathbf{k}-\mathbf{q}|^3}\,,
\end{align}
\begin{align}
&\langle S_\lambda(k,\tau_1)S_\lambda(k',\tau_2)\rangle'_{NG,h}\equiv{8^2(2\pi^2)}\tilde \fnlsq\int \frac{d^3q}{2\pi}  \left(e_\lambda^{ij}(k)q_iq_j\right)^2f(\tau_1,q,|\mathbf{k}-\mathbf{q}|)f(\tau_2,q,|\mathbf{k}-\mathbf{q}|)\nonumber\\&
\times\int \frac{d^3l}{2\pi}\left(\frac{{\cal P}_\Phi(|\mathbf{k}-\mathbf{q}|)}{|\mathbf{k}-\mathbf{q}|^3}\frac{{\cal P}_\Phi(l)}{l^3}\frac{{\cal P}_\Phi(|\mathbf{q}-\mathbf{l}|)}{|\mathbf{q}-\mathbf{l}|^3}+\frac{{\cal P}_\Phi(q)}{q^3}\frac{{\cal P}_\Phi(l)}{l^3}\frac{{\cal P}_\Phi(|\mathbf{k}-\mathbf{q}-\mathbf{l}|)}{|\mathbf{k}-\mathbf{q}-\mathbf{l}|^3}\right)\,,
\end{align}
and
\begin{align}
&\langle S_\lambda(k,\tau_1)S_\lambda(k',\tau_2)\rangle'_{NG,w}\equiv{8^2(2\pi^2)} \tilde \fnlsq\int \frac{d^3q}{2\pi}  e_\lambda^{ij}(k)q_iq_jf(\tau_1,q,|\mathbf{k}-\mathbf{q}|)\nonumber\\&
\times\int \frac{d^3l}{2\pi} e_\lambda^{ij}(k)l_il_j f(\tau_2,l,|\mathbf{k}-\mathbf{l}|)\Bigg(\frac{{\cal P}_\Phi(|\mathbf{k}-\mathbf{q}|)}{|\mathbf{k}-\mathbf{q}|^3}\frac{{\cal P}_\Phi(l)}{l^3}\frac{{\cal P}_\Phi(|\mathbf{q}-\mathbf{l}|)}{|\mathbf{q}-\mathbf{l}|^3}+\frac{{\cal P}_\Phi(q)}{q^3}\frac{{\cal P}_\Phi(l)}{l^3}\frac{{\cal P}_\Phi(|\mathbf{k}-\mathbf{q}-\mathbf{l}|)}{|\mathbf{k}-\mathbf{q}-\mathbf{l}|^3}\nonumber\\&+\frac{{\cal P}_\Phi(|\mathbf{k}-\mathbf{q}|)}{|\mathbf{k}-\mathbf{q}|^3}\frac{{\cal P}_\Phi(|\mathbf{k}-\mathbf{l}|)}{|\mathbf{k}-\mathbf{l}|^3}\frac{{\cal P}_\Phi(|\mathbf{k}-\mathbf{q}-\mathbf{l}|)}{|\mathbf{k}-\mathbf{q}-\mathbf{l}|^3}+\frac{{\cal P}_\Phi(|\mathbf{k}-\mathbf{l}|)}{|\mathbf{k}-\mathbf{l}|^3}\frac{{\cal P}_\Phi(q)}{q^3}\frac{{\cal P}_\Phi(|\mathbf{q}-\mathbf{l}|)}{|\mathbf{q}-\mathbf{l}|^3}\Bigg)\,.
\end{align}

To compute the power spectrum of induced GWs we sum over polarizations, explicitly
\begin{align}
{\cal P}_h=&\frac{k^3}{2\pi^2}\left(\langle h_+(k,\tau)h_+(k',\tau)\rangle'+\langle h_\times(k,\tau)h_\times(k',\tau)\rangle'\right)\nonumber\\&=\frac{k^3}{2\pi^2}\int_0^{k\tau} d(k\tau_1)\int_0^{k\tau} d(k\tau_1)\frac{1}{k^4}G(k\tau,k\tau_1)G(k\tau,k\tau_2)\nonumber\\&
\qquad\qquad\times\left(\langle S_+(k,\tau_1)S_+(k,\tau_2)\rangle+\langle S_\times(k,\tau_1)S_\times(k,\tau_2)\rangle\right)\,.
\end{align}
%\begin{align}
%P_h=&\,8\int dvdu\frac{d\varphi_k}{2\pi}\left(\frac{4v^2-(1-u^2+v^2)^2}{4u^2}\right){{\cal P}_\Phi(ku)}I(\tau,k,v,u)\nonumber\\\times\Bigg\{\left(\frac{4v^2-(1-u^2+v^2)^2}{4v^2}\right)&I(\tau,k,v,u)\left({{\cal P}_\Phi(kv)}+2F_{\rm NL}^2\int \frac{drds}{r^2s^2}\frac{d\varphi_l}{2\pi}{{\cal P}_\Phi(kvr)}{{\cal P}_\Phi(kvs)}\right)\nonumber\\&+4F_{\rm NL}^2\int \frac{drds}{s^2}\frac{d\varphi_l}{2\pi} I_2(\tau,l,|\mathbf{k}-\mathbf{l}|) e_\lambda^{ij}(k)l_il_j {{\cal P}_\Phi(kvr)}{{\cal P}_\Phi(kvs)}\Bigg\}
%\end{align}
As in Eq.~\eqref{eq:sl}, we divide ${\cal P}_h$ into three parts coming from different contributions, that is
\begin{align}
{\cal P}_h(k)={\cal P}_h^g(k)+{\cal P}_h^{NG,h}(k)+{\cal P}_h^{NG,w}(k)\,.
\end{align}
The different contributions are explicitly given by

\begin{align}
{\cal P}_h^g=8\int_0^\infty dv\int_{|1-v|}^{1+v}du\left(\frac{4v^2-(1-u^2+v^2)^2}{4uv}\right)^2\overline{I^2(\tau,k,v,u)}{{\cal P}_{\cal R}(ku)}{{\cal P}_{\cal R}(kv)}\,,
\end{align}
\begin{align}\label{eq:hybridexplicit}
{\cal P}_h^{NG,h}(k)=&16F_{\rm NL}^2\int_0^\infty dv\int_{|1-v|}^{1+v}du\left(\frac{4v^2-(1-u^2+v^2)^2}{4uv}\right)^2\overline{I^2(\tau,k,v,u)}{{\cal P}_{\cal R}(ku)} \nonumber\\&
\times q^3\int \frac{d^3l}{2\pi} \frac{{\cal P}_{\cal R}(l)}{l^3} \frac{{\cal P}_{\cal R}(|\mathbf{q}-\mathbf{l}|)}{|\mathbf{q}-\mathbf{l}|^3}\,,
\end{align}
and
\begin{align}\label{eq:walnutexplicit}
{\cal P}_h^{NG,w}(k)=&32F_{\rm NL}^2 \int_0^\infty dv\int_{|1-v|}^{1+v}du\left(\frac{4v^2-(1-u^2+v^2)^2}{4uv}\right)\frac{v^2}{u}{{\cal P}_{\cal R}(ku)}\nonumber\\&
\times \int_0^\infty d\tilde v\int_{|1-\tilde v|}^{1+\tilde v}d\tilde u\,\tilde u  \left(1-\frac{(1-\tilde u^2+\tilde v^2)^2}{4\tilde v^2}\right)\overline{I(\tau,k,v,u)I(\tau,k,\tilde v,\tilde u)}{{\cal P}_{\cal R}(k\tilde v)}\nonumber\\&
\times\frac{1}{\pi}\int_{-1}^{1} dx\frac{2x^2-1}{\sqrt{1-x^2}}\frac{{\cal P}_{\cal R}(kZ)}{Z^3}\,.
\end{align}
In the hybrid and walnut contributions we used the symmetries between $\mathbf{l}$ and $\mathbf{q}$ to simplify the expression to a single term. We also used the projections of the polarization tensors and the momenta that are given in App.~\ref{app:tensorpolarization} in which we choose $\mathbf{k}$ in the $z$-axis. We also introduced several definitions. We defined new variables as
\begin{align}
v\equiv\frac{q}{k}\quad,\quad u\equiv\frac{|\mathbf{k}-\mathbf{q}|}{k}\quad,\quad \tilde v\equiv\frac{l}{k}\quad,\quad \tilde u\equiv\frac{|\mathbf{k}-\mathbf{l}|}{k}\,.
\end{align}
We defined the kernel $I(\tau,k,x,y)$ by
\begin{align}\label{eq:kernelApp}
I(\tau,k,x,y)\equiv 2\left(\frac{2+b}{3+2b}\right)^2\int_0^{k\tau}d(k\tau_1)G(k\tau,k\tau_1)f(\tau_1,k,x,y)\,,
\end{align}
and redefined the non-Gaussian parameter to
\begin{align}
F_{\rm NL}\equiv\frac{2+b}{3+2b} \tilde \fnl\,,
\end{align}
since we related the Newtonian potential $\Phi_q$ with the primordial curvature perturbation ${\cal R}$\,, namely
\begin{align}
\Phi_p(k)=\frac{2+b}{3+2b}{\cal R}(k)\,.
\end{align}
In Eq.~\eqref{eq:walnutexplicit} we introduced the compact notation of $Z$ that is given by
\begin{align}
Z=\frac{|\mathbf{q}-\mathbf{l}|}{k}\,,
\end{align}
and that includes the azimuthal dependence of the walnut diagram, encoded in the variable $x$ by
\begin{align}
x=\cos(\varphi_q-\varphi_l)\,.
\end{align}
In terms of the new variables, $Z$ explicitly reads
\begin{align}
2Z^2=-1+u^2+v^2&+\tilde u^2+\tilde v^2-(u^2-v^2)(\tilde u^2-\tilde v^2)\nonumber\\&-x\sqrt{\left(4v^2-(1-u^2+v^2)^2\right)\left(4\tilde v^2-(1-\tilde u^2+\tilde v^2)^2\right)}\,.
\end{align}
It should be noted that in Eq.~\eqref{eq:hybridexplicit} we do not expand the integral over $\mathbf{l}$ since it can be reabsorbed into a redefinition of the primordial power spectrum including non-Gaussianities by
\begin{align}
{\cal P}_{\cal R}^{NL}(q)={\cal P}_{\cal R}(q)+F^2_{NL}q^3\int \frac{d^3l}{2\pi} \frac{{\cal P}_{\cal R}(l)}{l^3} \frac{{\cal P}_{\cal R}(|\mathbf{q}-\mathbf{l}|)}{|\mathbf{q}-\mathbf{l}|^3}\,,
\end{align}
which follows from the computation of $\langle\Phi(k)\Phi(k')\rangle$ using Eq.~\eqref{eq:FNLPHIExpansion}.

\section{Tensor polarizations\label{app:tensorpolarization}}
In this appendix we present the details of the tensor polarizations used in the main text and their projections. We start from the definition of the polarization tensor from the polarization vectors
\begin{align}
e^{+}_{ij}(\mathbf{k})&=\frac{1}{\sqrt{2}}\left[e_{i}(\mathbf{k})e_{j}(\mathbf{k})-\bar e_{i}(\mathbf{k})\bar e_{j}(\mathbf{k})\right]\,,\\
e^{\times}_{ij}(\mathbf{k})&=\frac{1}{\sqrt{2}}\left[e_{i}(\mathbf{k})\bar e_{j}(\mathbf{k})+\bar e_{i}(\mathbf{k}) e_{j}(\mathbf{k})\right]\,.
\end{align}
Using that $e_{i}(\mathbf{k})k^i=\bar e_{i}(\mathbf{k})k^i=0$ and $e_{i}(\mathbf{k})e^{i}(\mathbf{k})=\bar e_{i}(\mathbf{k})\bar e^{i}(\mathbf{k})=1$, we arrive at the usual properties of the polarization tensors, namely
\begin{align}
e^{+}_{ij}(\mathbf{k})e^{+ ij}(-\mathbf{k})&=1\quad,\quad e^{\times}_{ij}(\mathbf{k})e^{\times ij}(-\mathbf{k})=1\quad,\quad e^{+}_{ij}(\mathbf{k})e^{\times ij}(-\mathbf{k})=0\,,\nonumber\\
\delta_{ij}e^{+ij}(\mathbf{k})&=\delta_{ij}e^{\times ij}(\mathbf{k})=k_ie^{+ij}(\mathbf{k})=k_ie^{\times ij}(\mathbf{k})=0\,.
\end{align}

Let us start with the general case where the wave-vector of the tensor modes is given in spherical coordinates with respect to some fixed coordinate system by
\begin{align}
\mathbf{k}=k(\sin\theta_k\cos\varphi_k,\sin\theta_k\sin\varphi_k,\cos\theta_k)\,,
\end{align}
where  $\theta_k$ and $\varphi_k$ respectively are the polar and azimuthal angles. In this case, we have that an orthonormal set of polarization vectors is given by
\begin{align}
\mathbf{e}(\mathbf{k})&=(\cos\theta_k\cos\varphi_k,\cos\theta_k\sin\varphi_k,-\sin\theta_k)\\
\bar{\mathbf{e}}(\mathbf{k})&=(-\sin\varphi_k,\cos\varphi_k,0)\,.
\end{align}

Let us introduce a pair of wave-vectors of the scalar modes, say $\mathbf{q}$ and $\mathbf{l}$ which in spherical coordinates are given by
\begin{align}
\mathbf{q}=q(\sin\theta_{q}\cos\varphi_{q},\sin\theta_{q}\sin\varphi_{q},\cos\theta_{q})\,,
\end{align}
and
\begin{align}
\mathbf{l}=l(\sin\theta_{l}\cos\varphi_{l},\sin\theta_{l}\sin\varphi_{l},\cos\theta_{l})\,.
\end{align}
Note that since in the integral concerning the walnut diagram \eqref{eq:walnutexplicit} there is a dependence on $|\mathbf{q}-\mathbf{l}|$, $|\mathbf{k}-\mathbf{l}|$ and $|\mathbf{k}-\mathbf{q}|$ there is no possibility of choosing a reference system where the azimuthal angle dependence between all vectors disappears. Since the three vectors play a role in the integral, we have to take into account 3 angles. This is clear by looking at the explicit expressions of the modulus of the differences, which are given by
\begin{align}
|\mathbf{k}-\mathbf{q}|^2&=k^2+q^2-2kq\left(\cos\theta_k\cos\theta_{q}+\cos(\varphi_k-\varphi_q)\sin\theta_k\sin\theta_{q}\right)\,,\\
|\mathbf{k}-\mathbf{l}|^2&=k^2+l^2-2kl\left(\cos\theta_k\cos\theta_{l}+\cos(\varphi_k-\varphi_l)\sin\theta_k\sin\theta_{l}\right)\,,\\
|\mathbf{q}-\mathbf{l}|^2&=q^2+l^2-2lq\left(\cos\theta_q\cos\theta_{l}+\cos(\varphi_q-\varphi_l)\sin\theta_q\sin\theta_{l}\right)\,.\\
%|\mathbf{k}-\mathbf{q}-\mathbf{l}|^2&=k^2+|\mathbf{q}-\mathbf{l}|^2-2k\left(\cos\theta_k(l\cos\theta_l-q\cos\theta_q)+\sin\theta_k(l\cos(\varphi_k-\varphi_l)\sin\theta_l-q\cos(\varphi_k-\varphi_q)\sin\theta_q\right)\,.
\end{align}
We also have that
\begin{align}
|\mathbf{k}-\mathbf{q}|^2+|\mathbf{k}-\mathbf{l}|^2-|\mathbf{q}-\mathbf{l}|^2-|\mathbf{k}-\mathbf{q}-\mathbf{l}|^2-k^2+q^2+l^2=0
\end{align}
so we can write $|\mathbf{k}-\mathbf{q}-\mathbf{l}|^2$ in terms of the others.

Due to the symmetry between $\mathbf{l}$ and $\mathbf{q}$, we find most convenient to choose $\mathbf{k}$ in the z-axis, namely we set
\begin{align}
\theta_k=0\quad,\quad \varphi_k=0\,.
\end{align}
Then we have that
\begin{align}
|\mathbf{k}-\mathbf{q}|^2&=k^2+q^2-2kq\cos\theta_k\,,\\
|\mathbf{k}-\mathbf{l}|^2&=k^2+l^2-2kl\cos\theta_l\,,\\
|\mathbf{q}-\mathbf{l}|^2&=q^2+l^2-2lq\left(\cos\theta_{q}\cos\theta_{l}+\cos(\varphi_q-\varphi_l)\sin\theta_{q}\sin\theta_{l}\right)\,,
\end{align}
where the azimuthal dependence is only in $|\mathbf{q}-\mathbf{l}|$\,. In such coordinate system the different projections yield
\begin{align}
e^{+}_{ij}(\mathbf{k})q^iq^j&=\frac{1}{\sqrt{2}}q^2\sin^2\theta_{k}\cos(2\varphi_q)\quad,\quad
e^{\times}_{ij}(\mathbf{k})q^iq^j=\frac{1}{\sqrt{2}}q^2\sin^2\theta_{k}\sin(2\varphi_q)\\
e^{+}_{ij}(\mathbf{k})l^il^j&=\frac{1}{\sqrt{2}}l^2\sin^2\theta_{l}\cos(2\varphi_l)\quad,\quad
e^{\times}_{ij}(\mathbf{k})l^il^j=\frac{1}{\sqrt{2}}l^2\sin^2\theta_{l}\sin(2\varphi_l)\,.
\end{align}
Regarding the combinations that enter the integrals in the power spectrum of induced GWs we obtain
\begin{align}
\left(e^{+}_{ij}(\mathbf{k})q^iq^j\right)^2+\left(e^{\times}_{ij}(\mathbf{k})q^iq^j\right)^2=\frac{q^4}{2}\sin^4\theta_{k}\,,
\end{align}
and
\begin{align}
e^{+}_{ij}(\mathbf{k})q^iq^je^{+}_{ij}(\mathbf{k})l^il^j+e^{\times}_{ij}(\mathbf{k})q^iq^je^{\times}_{ij}(\mathbf{k})l^il^j=\frac{q^2l^2}{2}\sin^2\theta_{k}\sin^2\theta_{l}\cos(2(\varphi_q-\varphi_l))\,.
\end{align}
It should be noted that for the computations of the hybrid diagram it is most convenient to choose $q$ in the $z$-axis, that is $\theta_q=\varphi_q=0$. This is because in the hybrid diagram integral only the angle between $\mathbf{l}$ and $\mathbf{q}$ matters for the integral over $d^3l$. Thus, referring all quantities with respect to $\mathbf{q}$ is the most suitable choice for calculations.

\section{Induced GWs from the connected 4-point function\label{app:walnut}}

In this appendix we estimate the contribution from the walnut diagram which is the leading order term in $\fnl$ from the connected 4-point function. Thus, we focus our attention on how to estimate the general integral \eqref{eq:walnutexplicit}. We shall focus for simplicity in a power-law curvature power-spectrum. The main problematic term is the integral over the azimuthal angle, which is denoted by $x=\cos(\varphi_q-\varphi_l)$ in \eqref{eq:walnutexplicit}. Once we find an approximate expression for the integral over $x$ we shall proceed equivalently as the case for the other diagrams. 

\subsection{The hybrid diagram integral}

We start by noting that in the hybrid non-Gaussian case, we have a similar integral which is given by
\begin{align}
{\cal I}_h(q)=q^3\int \frac{d^3l}{2\pi} \frac{{\cal P}_{\cal R}(l)}{l^3} \frac{{\cal P}_{\cal R}(|\mathbf{q}-\mathbf{l}|)}{|\mathbf{q}-\mathbf{l}|^3}\,.
\end{align}
The smart way to do this integral is to choose $\mathbf{q}$ as the $z$-axis and then use $|\mathbf{q}-\mathbf{l}|$ as angular variable. In this case, we have
\begin{align}\label{eq:smartx}
{\cal I}_h(q)=\int_0^\infty d\hat v \int_{|1-\hat v|}^{1+\hat v} du  \frac{1}{\hat v^2\hat u^2} {\cal P}_{\cal R}(q\hat v){\cal P}_{\cal R}(q\hat u)\,.
\end{align}
where
\begin{align}
\hat v\equiv \frac{l}{q}\quad,\quad \hat u\equiv\frac{|\mathbf{q}-\mathbf{l}|}{q}\,.
\end{align}
Now, let us do the integral by brute force. Let us choose a vector $\mathbf{k}$ as reference, e.g. as $z$-axis. Then the integral in spherical coordinates becomes 
\begin{align}\label{eq:complicatedx}
{\cal I}_h(q)=\frac{v^3}{\pi}\int_0^\infty d\tilde v \int_{|1-\tilde v|}^{1+\tilde v} d\tilde u  \frac{\tilde u}{\tilde v^2} {\cal P}_{\cal R}(k\tilde v)\int_{-1}^{1} dx\frac{1}{\sqrt{1-x^2}}\frac{{\cal P}_{\cal R}(kZ)}{Z^3}\,,
\end{align}
where we defined
\begin{align}
v\equiv \frac{q}{k}\quad,\quad  u\equiv\frac{|\mathbf{k}-\mathbf{q}|}{k}\quad,\quad \tilde v\equiv \frac{l}{k}\quad,\quad \tilde u\equiv\frac{|\mathbf{q}-\mathbf{l}|}{k}\,,
\end{align}
and
\begin{align}
2Z^2=-1+u^2+v^2&+\tilde u^2+\tilde v^2-(u^2-v^2)(\tilde u^2-\tilde v^2)\nonumber\\&-x\sqrt{\left(4v^2-(1-u^2+v^2)^2\right)\left(4\tilde v^2-(1-\tilde u^2+\tilde v^2)^2\right)}\,.
\end{align}

We may find the value of the integral over $x$ by comparing with the two approaches. Let us consider that 
\begin{align}
{\cal P}_{\cal R}(k)=\left(\frac{k}{k_p}\right)^{3+2\alpha}\,,
\end{align}
with an appropriate UV or IR cut-off at $k_p$ depending on the value of $\alpha$. The $x$ integral then can be written as
\begin{align}
{\cal I}_{h,x}=\frac{1}{\pi}\int_{-1}^{1} dx\frac{1}{\sqrt{1-x^2}}\frac{{\cal P}_{\cal R}(kZ)}{Z^3}=\left(\frac{k}{k_p}\right)^{3+2\alpha}\frac{1}{\pi}\int_{-1}^{1} dx\frac{(A-Bx)^{\alpha}}{\sqrt{1-x^2}}\,,
\end{align}
where
\begin{align}
{2}A&=-1+u^2+v^2+\tilde u^2+\tilde v^2-(u^2-v^2)(\tilde u^2-\tilde v^2)\\
{2}B&=\sqrt{\left(4v^2-(1-u^2+v^2)^2\right)\left(4\tilde v^2-(1-\tilde u^2+\tilde v^2)^2\right)}\,.
\end{align}
We did not consider the effects of the cut-off on $Z$ at this stage by analogy with previous calculation where only the cut-off on $\tilde v$ and $v$ played a role. There is an analytical expression for such integral which reads
\begin{align}\label{eq:xanaliticalh}
{\cal I}_{h,x}=A^\alpha\left(\frac{k}{k_p}\right)^{3+2\alpha} \, _2F_1\left(\frac{1-\alpha}{2},-\frac{\alpha}{2};1;\frac{B^2}{A^2}\right)\,.
\end{align}
We shall study the two limiting cases of the IR and UV parts of the spectrum taking into account that we doing the integrals over $\tilde v$ and $\tilde u$ before $v$ and $u$. Although in the end we may have that due to the cut-off $v\sim\tilde v\sim v_p$ we shall assume, for the sake of an analytical approximation, that there is always a hierarchy between $\tilde v$ and $v$ to be specified later.

\subsubsection{IR limit}
Here we look at the $\tilde u\sim \tilde v\gg1$ limit. First, using $q$ as $z$-axis, we find using Eq.~\eqref{eq:smartx} that
\begin{align}
{\cal I}_h(q)\approx \int_0^{\hat v_p} d\hat v \frac{1}{\hat v^4} {\cal P}_{\cal R}^2(q\hat v)\,.
\end{align}
On the other hand, by brute fore in Eq.~\eqref{eq:complicatedx} we arrive at
\begin{align}
{\cal I}_h(q)=\frac{v^3}{\pi}\int_0^\infty d\tilde v \int_{|1-\tilde v|}^{1+\tilde v} d\tilde u  \frac{1}{\tilde v} {\cal P}_{\cal R}(k\tilde v)\int_{-1}^{1} dx\frac{1}{\sqrt{1-x^2}}\frac{{\cal P}_{\cal R}(kZ)}{Z^3}\,,
\end{align}
where we introduced a UV cut-off at $\hat v_p=k_p/q$ and $v_p=k_p/k$.  By comparing the two and noting the relation between $\hat v$ and $\tilde v$, that is $\hat v=\tilde v/v$ we conclude that
\begin{align}\label{eq:approx1}
{\cal I}_{h,x}\approx \frac{1}{\tilde v^3}{\cal P}_{\cal R}(k\tilde v)\,.
\end{align}

We can recover the same expression from the analytical formula \eqref{eq:xanaliticalh} by using that, in the IR limit, 
\begin{align}
A\approx \tilde v^2\quad,\quad B\approx 2v\tilde v\,,
\end{align}
where in the last step we also used $u\sim v\gg1$. Then we find that
\begin{align}
\frac{B}{A}\approx \frac{2 v}{\tilde v}\,.
\end{align}
In the current order of integration we have that $\tilde v>v$ and so $B/A\ll1$. Using this limit in Eq.~\eqref{eq:xanaliticalh} we recover Eq.~\eqref{eq:approx1}.

\subsubsection{UV limit}
Let us turn to the $\tilde u\sim1$, $ \tilde v\ll 1$ limit. Again, using $q$ as $z$-axis, we find using Eq.~\eqref{eq:smartx} that
\begin{align}
{\cal I}_h(q)\approx 2{\cal P}_{\cal R}(q)\int_{\hat v_p}^{\infty} d\hat v  \frac{1}{\hat v} {\cal P}_{\cal R}(q\hat v)\,.
\end{align}
Now, from Eq.~\eqref{eq:complicatedx} we have
\begin{align}
{\cal I}_h(q)=2\frac{v^3}{\pi}\int_0^\infty d\tilde v \frac{1}{\tilde v} {\cal P}_{\cal R}(k\tilde v)\int_{-1}^{1} dx\frac{1}{\sqrt{1-x^2}}\frac{{\cal P}_{\cal R}(kZ)}{Z^3}\,,
\end{align}
where we introduced a IR cut-off at $\hat v_p=k_v/q$ and $v_p=k_p/k$.  By comparing the two and noting the relation between $\hat v$ and $\tilde v$, that is $\hat v=\tilde v/v$ we conclude that
\begin{align}\label{eq:approx2}
{\cal I}_{h,x}\approx \frac{1}{v^3}{\cal P}_{\cal R}(kv)\,.
\end{align}

We can recover the same expression from the analytical formula \eqref{eq:xanaliticalh} by using that, in the UV limit, 
\begin{align}
A\approx v^2\quad,\quad B\approx 2v\tilde v\,,
\end{align}
where we used $u\sim1$ and $v\ll1$. This time, we find that
\begin{align}
\frac{B}{A}\approx \frac{2 \tilde v}{ v}\,.
\end{align}
In the current order of integration we have that $\tilde v<v$ and so $B/A\ll1$. Using this limit in Eq.~\eqref{eq:xanaliticalh} we recover Eq.~\eqref{eq:approx2}. Since we developed a consistent way to treat the $x$ integral in the hybrid diagram integral, we shall apply the same mechanism to the walnut diagram integral.

\subsection{The walnut diagram integral}

In the walnut case we instead have an integral of the type
\begin{align}
{\cal I}_{w,x}=\int_{-1}^{1} dx\frac{2x^2-1}{\sqrt{1-x^2}}\frac{{\cal P}_{\cal R}(kZ)}{Z^3}=\left(\frac{k}{k_p}\right)^{3+2\alpha}\frac{1}{\pi}\int_{-1}^{1} dx\frac{2x^2-1}{\sqrt{1-x^2}}(A-Bx)^{\alpha}\,.
\end{align}
We see that the main difference from the hybrid case is the $x$ dependence in the numerator. This may introduce numerical differences of the coefficients but it should not alter the dependence in $k$, $v$ and $\tilde v$ derived in Eqs.~\eqref{eq:approx1} and \eqref{eq:approx2}. From this we conclude that the walnut diagram does not change the IR or UV behaviors of the induced GW spectrum. However, the weight of the diagram depends on the form of ${\cal P}_{\cal R}(k)$. For example, if ${\cal P}_{\cal R}(k)\propto k^3$ the walnut diagram does not contribute at all as
\begin{align}
\int_{-1}^{1} dx\frac{2x^2-1}{\sqrt{1-x^2}}=\int_0^{2\pi}d\varphi \cos \varphi=0\,.
\end{align}
Nevertheless, regarding the $k$ dependence of the induced GW spectrum IR and UV tails and the relation to the $\fnl$ in single-field inflationary models is independent of the weight of the walnut diagram. We shall proceed to check these intuitive statements.

We again neglect the UV or IR cut-off on $Z$ and use the following analytic expression:
\begin{align}
{\cal I}_{w,x}=A^\alpha\left(\frac{k}{k_p}\right)^{3+2\alpha}  \left(\,
   _3F_2\left(\frac{3}{2},\frac{1-\alpha}{2},-\frac{\alpha}{2};\frac{1}{2},2;\frac{B^2}{A
   ^2}\right)-\, _2F_1\left(\frac{1-\alpha}{2},-\frac{\alpha}{2};1;\frac{B^2}{A^2}\right)\right)\,.
\end{align}
In the limit where $B\ll A$ we have that
\begin{align}\label{eq:approxwalnutx}
{\cal I}_{w,x}\approx \frac{\alpha(\alpha-1)}{8}\frac{B^2}{A^2}A^\alpha\left(\frac{k}{k_p}\right)^{3+2\alpha} \,.
\end{align}
We see that the integral over $x$ in the walnut diagram is zero at leading order in $B/A\ll1$. Thus, we took into account the next to leading order. We also find that for $\alpha=\{0,1\}$ the walnut contribution is zero. The former corresponds to ${\cal P}_{\cal R}\propto k^3$ while the latter to  ${\cal P}_{\cal R}\propto k^5$. Also, the overall $k$ dependence is the same as in the hybrid case, since at the end we have to consider first that $\tilde v\sim v_p$ and later $v\sim v_p$.

\subsection{The walnut contribution to the induced GWs}

Once we know the behavior of the $x$ integral in the IR and UV limits, we are ready to estimate the amplitude and $k$ dependence of the induced GWs from the walnut diagram contribution. As before, we shall divide the analysis into the IR and the UV limits. We also compare the contribution of the walnut with respect to the hybrid.

\subsubsection{IR limit}
Proceeding as in the previous section, we focus in the regime where $u\sim v\gg1$. Expanding the integrand in Eq.~\eqref{eq:walnutexplicit} we arrive at
\begin{align}
\Omega_{\rm GW}^w(k\ll k_p)\approx {\cal A}_{\cal R}^3F_{\rm NL}^2\frac{6\alpha(\alpha-1)}{(5+2\alpha)(1+4\alpha)}\left(\frac{k}{k_p}\right)^{3} \ln^2\left(\frac{k}{k_p}\right)\,,
\end{align}
A similar calculation for the hybrid contribution \eqref{eq:hybridexplicit} yields
\begin{align}
\Omega_{\rm GW}^h(k\ll k_p)\approx {\cal A}_{\cal R}^3F_{\rm NL}^2\frac{24}{(3+2\alpha)(3+4\alpha)}\left(\frac{k}{k_p}\right)^{3} \ln^2\left(\frac{k}{k_*}\right)\,.
\end{align}
In the main text we considered that in the IR ${\cal P}_{\cal R}\propto k^4$, that is $\alpha=1/2$. In this case, we find
\begin{align}
\frac{\Omega_{\rm GW}^h(k\ll k_p)}{\Omega_{\rm GW}^w(k\ll k_p)}\sim -15\,.
\end{align}
Thus we can safely neglect the walnut contribution in the IR limit. It is interesting to note that the walnut diagram yields a negative contribution for $0<\alpha<1$, although it is negligible compared to the hybrid one.

\subsubsection{UV limit}
Now we study the opposite limit, that is $u\sim 1$ and $v\ll1$. As in Sec.~\ref{subsec:GWpowerlaw} we distinguish the cases where $3+2\alpha<-4$ and $3+2\alpha>-4$. In the former, the integral does not converge and the dominant contribution arises at the cut-off. In the latter, the integral converges and the estimation of the amplitude becomes a numerical issue.

First, for the case that $3+2\alpha<-4$ we find that Eq.~\eqref{eq:walnutexplicit} in the UV limit yields
\begin{align}
\Omega_{\rm GW}^w(k\gg k_p)\approx {\cal A}_{\cal R}^3F_{\rm NL}^2\frac{32\alpha(\alpha-1)}{3(3+2\alpha)(7+2\alpha)}\left(\frac{k}{k_p}\right)^{1-2\alpha}\,.
\end{align}
The hybrid contribution \eqref{eq:hybridexplicit} is now given by
\begin{align}
\Omega_{\rm GW}^h(k\gg k_p)\approx {\cal A}_{\cal R}^3F_{\rm NL}^2\frac{16}{3(3+2\alpha)(7+2\alpha)}\left(\frac{k}{k_p}\right)^{1-2\alpha}\,.
\end{align}
For $\alpha=-4$, i.e. ${\cal P}_{\cal R}\propto k^{-5}$, we obtain
\begin{align}
\frac{\Omega_{\rm GW}^h(k\gg k_p)}{\Omega_{\rm GW}^w(k\gg k_p)}\sim \frac{1}{40}\,.
\end{align}
Thus, in the UV limit the walnut diagram is dominant over the hybrid one. However, this only changes the amplitude of the GW spectrum and not the UV slope. 

Second, the case when $0>3+2\alpha>-4$ is more complicated as the integral converges everywhere. Then, the dominant contribution comes from the divergence in the kernel. However, estimating the amplitude analytically is a subtle task. In this case, numerical methods are more suitable. Due to the involved numerical computations in the case of the walnut we postpone this issue for future work. Nevertheless, assuming that the ratio between the hybrid and walnut diagrams is a monotonic function of $\alpha$ times $\alpha(1-\alpha)$, it is reasonable to expect that for $0>3+2\alpha>-4$ both diagrams would have similar amplitude.

\section{The kernel in general backgrounds \label{app:kernelgeneral}}

In this appendix we present the general formulas derived in Refs.\cite{Domenech:2019quo,Domenech:2020kqm} for the kernel \eqref{eq:kernelApp} when the induced GWs are generated in a universe dominated by a perfect fluid with equation of state $w=p/\rho$ and sound speed $c_s$. The expressions of Refs.\cite{Domenech:2019quo,Domenech:2020kqm} are derived for the cases where $c_s^2=w>0$ and $c_s^2=1$ with $w>-1/3$. Therefore, we shall only consider such situations.

We have that in the subhorizon limit, i.e. for scales that entered the horizon before the universe is reheated, the kernel reads
\begin{align}\label{eq:kernel33}
I(u,v,b,x\gg1)=x^{-(b+1/2)}\left(C_{1,b}J_{b+1/2}(x)+C_{2,b}Y_{b+1/2}(x)\right)\,,
\end{align}
where $J_{b+1/2}(x)$ and $Y_{b+1/2}(x)$ are Bessel functions, we have introduced for convenience
\begin{align}
b\equiv\frac{1-3w}{1+3w}\,,
\end{align}
and the time independent coefficients are given by
\begin{align}\label{eq:c1b}
C_{1,b}=-2^{1+2b}\pi\frac{2+b}{3+2b}&\Gamma^2[b+3/2](c^2_s uv)^{-b-1/2}{\cal I}_Y(u,v,b,x\gg1)
\end{align}
and
\begin{align}\label{eq:c2b}
C_{2,b}=2^{1+2b}\pi\frac{2+b}{3+2b}&\Gamma^2[b+3/2](c^2_s uv)^{-b-1/2}{\cal I}_J(u,v,b,x\gg1)\,,
\end{align}
where
\begin{align}\label{eq:IJ}
{\cal I}_{J}
(u,v,b,x\gg1)=(c_s^2 uv)^{b-1/2}\frac{\left(1-y^2\right)^{b/2}}{\sqrt{2\pi}}\left(\mathsf{P}^{-b}_{b}(y)+\frac{2+b}{1+b}\mathsf{P}^{-b}_{b+2}(y)\right)\Theta(u+v-c_s^{-2})\,,
\end{align}
and
\begin{align}\label{eq:IY}
{\cal I}_Y(u,v,b,x\gg1)=&-4(c_s^2 uv)^{b-1/2}\frac{\left(1-y^2\right)^{b/2}}{\left(2\pi\right)^{3/2}}\left(
	\mathsf{Q}^{-b}_{b}(y)+\frac{2+b}{1+b}\mathsf{Q}^{-b}_{b+2}(y)\right)\Theta(u+v-c_s^{-2})\nonumber\\&
	-4(c^2_s uv)^{b-1/2}\frac{\left(y^2-1\right)^{b/2}}{\left(2\pi\right)^{3/2}}\left(
	{\cal Q}^{-b}_{b}(-y)+2\frac{2+b}{1+b}{\cal Q}^{-b}_{b+2}(-y)\right)\Theta(c_s^{-2}-u-v)\,.
\end{align}
We also introduced a new variable for the sake of simplicity which is given by
\begin{align}\label{eq:y2}
y\equiv1-\frac{c_s^{-2}-(u-v)^2}{2uv}\,.
\end{align}
In Eqs.~\eqref{eq:IJ} and \eqref{eq:IY}, $\mathsf{P}^\mu_\nu(y)$ and $\mathsf{Q}^\mu_\nu(y)$ are Legendre functions on the cut (or Ferrer's functions) which are valid for $|y|<1$. Also, ${\cal Q}^\mu_\nu(y)$ is the associated Legendre function of the second kind and is valid for $|y|>1$. Their explicit expression in terms of hypergeometric functions is given in the next subsection.

The averaged kernel squared is given by
\begin{align}\label{eq:kernelaverage}
&\overline{I^2(u,v,b,x\gg1)}=x^{-2(1+b)}\,2^{1+4b}\left(\frac{2+b}{3+2b}\right)^2\Gamma^4[b+3/2](c_s^2uv)^{-2}\left|1-y^2\right|^b\nonumber\\&\times\Bigg\{\left[\left(\mathsf{P}^{-b}_{b}(y)+\frac{2+b}{1+b}\mathsf{P}^{-b}_{b+2}(y)\right)^2+\frac{4}{\pi^2}\left(
	\mathsf{Q}^{-b}_{b}(y)+\frac{2+b}{1+b}\mathsf{Q}^{-b}_{b+2}(y)\right)^2\right]\Theta(u+v-c_s^{-2})\nonumber\\&
	\qquad\qquad+\frac{4}{\pi^2}\left(
	{\cal Q}^{-b}_{b}(-y)+2\frac{2+b}{1+b}{\cal Q}^{-b}_{b+2}(-y)\right)^2\Theta(c_s^{-2}-u-v)\Bigg\}\,.
\end{align}

In the walnut diagram integral \eqref{eq:walnutexplicit} we have
\begin{align}\label{eq:kernelaverage2}
&\overline{I(u,v,b,x\gg1)I(\tilde u,\tilde v,b,x\gg1)}=\nonumber\\&x^{-2(1+b)}\,2^{1+4b}\left(\frac{2+b}{3+2b}\right)^2\Gamma^4[b+3/2](c_s^4uv\tilde u\tilde v)^{-1}\left|1-y^2\right|^{b/2}\left|1-\tilde y^2\right|^{b/2}\nonumber\\&\times\Bigg\{\left(\mathsf{P}^{-b}_{b}(y)+\frac{2+b}{1+b}\mathsf{P}^{-b}_{b+2}(y)\right)\left(\mathsf{P}^{-b}_{b}(\tilde y)+\frac{2+b}{1+b}\mathsf{P}^{-b}_{b+2}(\tilde y)\right)\Theta(u+v-c_s^{-2})\Theta(\tilde u+\tilde v-c_s^{-2})\nonumber\\&+\frac{4}{\pi^2}\Bigg[\left(
	\mathsf{Q}^{-b}_{b}(y)+\frac{2+b}{1+b}\mathsf{Q}^{-b}_{b+2}(y)\right)\Theta(u+v-c_s^{-2})+\left(
	{\cal Q}^{-b}_{b}(-y)+2\frac{2+b}{1+b}{\cal Q}^{-b}_{b+2}(-y)\right)\Theta(c_s^{-2}-u-v)\Bigg]\nonumber\\&
	\times\Bigg[\left(
	\mathsf{Q}^{-b}_{b}(\tilde y)+\frac{2+b}{1+b}\mathsf{Q}^{-b}_{b+2}(\tilde y)\right)\Theta(\tilde u+\tilde v-c_s^{-2})+\left(
	{\cal Q}^{-b}_{b}(-\tilde y)+2\frac{2+b}{1+b}{\cal Q}^{-b}_{b+2}(-\tilde y)\right)\Theta(c_s^{-2}-\tilde u-\tilde v)\Bigg]\Bigg\}\,.
\end{align}
where
\begin{align}\label{eq:y3}
\tilde y\equiv1-\frac{c_s^{-2}-(\tilde u-\tilde v)^2}{2\tilde u\tilde v}\,.
\end{align}

\subsection{Definition and asymptotics of Legendre functions}
Here we write down the useful expressions for the Legendre function which can be found in Ref.~\cite{NIST:DLMF}. The definitions of the Legendre functions of the first and second kind in terms of hypergeometric functions are given by
\begin{align}
\mathsf{P}^{\mu}_{\nu}\left(x\right)=\left(\frac{1+x}{1-x}\right)^{\mu/2}%
\mathbf{F}\left(\nu+1,-\nu;1-\mu;\tfrac{1}{2}-\tfrac{1}{2}x\right)\,,
\end{align}
\begin{align}
\mathsf{Q}^{\mu}_{\nu}\left(x\right)=&\frac{\pi}{2\sin\left(\mu\pi\right)}\Bigg%
(\cos\left(\mu\pi\right)\left(\frac{1+x}{1-x}\right)^{\mu/2}\mathbf{F}\left(%
\nu+1,-\nu;1-\mu;\tfrac{1}{2}-\tfrac{1}{2}x\right)\nonumber\\&\qquad\qquad\qquad-\frac{\Gamma\left(\nu+\mu+1%
\right)}{\Gamma\left(\nu-\mu+1\right)}\left(\frac{1-x}{1+x}\right)^{\mu/2}%
\mathbf{F}\left(\nu+1,-\nu;1+\mu;\tfrac{1}{2}-\tfrac{1}{2}x\right)\Bigg)\,,
\end{align}
\begin{align}
{\cal Q}^{\mu}_{\nu}\left(x\right)=&%
\frac{\pi}{2\sin\left(\mu\pi\right)\Gamma\left(\nu+\mu+1\right)}\Bigg(\frac{(x+1)^{\mu/2}}{(x-1)^{\mu/2}}\mathbf{F}\left%
(\nu+1,-\nu;1-\mu;\tfrac{1}{2}-\tfrac{1}{2}x\right)\nonumber\\&\qquad\qquad\qquad-\frac{\Gamma\left(\nu+\mu+1\right)(x-1)^{\mu/2}}{%
\Gamma\left(\nu-\mu+1\right)(x+1)^{\mu/2}}\mathbf{F}\left(\nu+1,-\nu;\mu+1;%
\tfrac{1}{2}-\tfrac{1}{2}x\right)\Bigg)\,.
\end{align}
Note that $\mathsf{P}^{\mu}_{\nu}(x)$ and  $\mathsf{Q}^{\mu}_{\nu}(x)$ are also referred to as Ferrer's functions and they are valid for $|x|<1$. ${\cal Q}^{\mu}_{\nu}(x)$ known as Olver's function which is real valuated for $|x|>1$.

In a radiation dominated universe we have that $b=0$ ($w=1/3$) and the Legendre functions are given by
\begin{align}
\mathsf{P}_0^0(x)&=1\quad,\quad \mathsf{P}_2^0(x)=\frac{1}{2}\left(3x^2-1\right)\\
\mathsf{Q}_0^0(x)&=\frac{1}{2}\ln\left(\frac{1+x}{1-x}\right)\quad,\quad \mathsf{Q}_2^0(x)=\frac{1}{4}\left(3x^2-1\right)\ln\left(\frac{1+x}{1-x}\right)-\frac{3}{2}x\,,
\end{align}
and
\begin{align}
%{P}_0^0(x)&=1\quad,\quad {P}_2^0(x)=\frac{1}{2}\left(3x^2-1\right)\\
{\cal Q}_0^0(x)=\frac{1}{2}\ln\left(\frac{x+1}{x-1}\right)\quad,\quad {\cal Q}_2^0(x)=\frac{1}{8}\left(3x^2-1\right)\ln\left(\frac{x+1}{x-1}\right)-\frac{3}{4}x\,.
\end{align}

We now present useful asymptotic behavior of the Legendre functions near the singular points. 

\paragraph{For $x\to 1^{-}$:}
For the Ferrer's function of the first kind we have that
\begin{align}
\mathsf{P}^{\mu}_{\nu}\left(x\right)\sim\frac{1}{\Gamma\left(1-\mu\right)}%
\left(\frac{2}{1-x}\right)^{\mu/2}.
\end{align}
For the Ferrer's function of the second kind, assuming $\mu>0$, we find
\begin{align}
\mathsf{Q}^{-\mu}_{\nu}\left(x\right)\sim\frac{\Gamma\left(\mu\right)\Gamma%
\left(\nu-\mu+1\right)}{2\Gamma\left(\nu+\mu+1\right)}\left(\frac{2}{1-x}%
\right)^{\mu/2},
\end{align}
and
\begin{align}
\mathsf{Q}^{\mu}_{\nu}\left(x\right)\sim\frac{1}{2}\cos\left(\mu\pi\right)%
\Gamma\left(\mu\right)\left(\frac{2}{1-x}\right)^{\mu/2}.
\end{align}

\paragraph{For $x\to 1^{+}$:} This case belongs to the Olver's function of the second kind. We find that
\begin{align}
{\cal Q}^{\mu}_{\nu}\left(x\right)\sim\frac{\Gamma\left(\mu\right)}{2%
\Gamma\left(\nu+\mu+1\right)}\left(\frac{2}{x-1}\right)^{\mu/2}\,.
\end{align}

\paragraph{For $x\to \infty$:} This case also concerns the Olver's function of the second kind. We obtain
\begin{align}
{\cal Q}^{\mu}_{\nu}\left(x\right)\sim\frac{\pi^{1/2}}{\Gamma\left(\nu+%
\frac{3}{2}\right)(2x)^{\nu+1}}\,.
\end{align}

\bibliographystyle{jhep}
\bibliography{bibliofull.bib}

\end{document}